\documentclass{revtex4-2}
\usepackage{amsmath}
\usepackage{amsfonts,euler,url,graphicx,subfigure}

\begin{document}

\title{Symmetry-breaking transitions in quiescent and moving solitons in
fractional couplers}
\author{Dmiry V. Strunin$^{1}$ and Boris A. Malomed$^{2,3}$}
\address{$^{1}$School of Mathematics, Physics and Computing,
University of Southern Queensland, Toowoomba, Queensland 4350,
Australia\\
$^{2}$Department of Physical Electronics, School of Electrical Engineering,
Faculty of Engineering, and Center for Light-Matter Interaction, Tel Aviv
University, P.O.B. 39040, Tel Aviv, Israel\\
$^{3}$Instituto de Alta Investigaci\'{o}n, Universidad de Tarapac\'{a},
Casilla 7D, Arica, Chile}

\begin{abstract}
We consider phase transitions, in the form of spontaneous symmetry breaking
(SSB) bifurcations of solitons, in dual-core couplers with fractional
diffraction and cubic self-focusing acting in each core, characterized by L%
\'{e}vy index $\alpha $. The system represents linearly-coupled optical
waveguides with the fractional paraxial diffraction or group-velocity
dispersion (the latter system was used in a recent experiment, which
demonstrated the first observation of the wave propagation in an effectively
fractional setup). By dint of numerical computations and variational
approximation (VA), we identify the SSB in the fractional coupler as the
bifurcation of the subcritical type (i.e., the symmetry-breaking phase
transition of the first kind), whose subcriticality becomes stronger with
the increase of fractionality $2-\alpha $, in comparison with very weak
subcriticality in the case of the non-fractional diffraction, $\alpha =2$.
In the Cauchy limit of $\alpha \rightarrow 1$, it carries over into the
\textit{extreme} subcritical bifurcation, manifesting backward-going
branches of asymmetric solitons which never turn forward. The analysis of
the SSB bifurcation is extended for moving (tilted) solitons, which is a
nontrivial problem because the fractional diffraction does not admit
Galilean invariance. Collisions between moving solitons are studied too,
featuring a two-soliton symmetry-breaking effect and merger of the solitons.
\end{abstract}

\maketitle

\section{Introduction}

The Schr\"{o}dinger equation with the kinetic-energy operator represented by
a fractional derivative was first derived by Laskin as a generalization of
canonical quantum mechanics \cite{Lask1,Baleanu,Lask2}. The scaled form of
the one-dimensional fractional Schr\"{o}dinger equation in the free space,
with the \textit{L\'{e}vy index} (LI) $\alpha $ (so named by Mandelbrot \cite%
{Mandelbrot}) is%
\begin{equation}
i\frac{\partial \psi }{\partial t}=\frac{1}{2}\left( -\frac{\partial ^{2}}{%
\partial x^{2}}\right) ^{\alpha /2}\psi ,  \label{fractional}
\end{equation}%
where the fractional operator\ is defined as the \textit{Riesz derivative}
\cite{Riesz1,Riesz2},%
\begin{equation}
\left( -\frac{\partial ^{2}}{\partial x^{2}}\right) ^{\alpha /2}\psi (x)=%
\frac{1}{2\pi }\int_{-\infty }^{+\infty }dp|p|^{\alpha }\int_{-\infty
}^{+\infty }dx^{\prime }e^{ip(x-x^{\prime })}\psi (x^{\prime }).
\label{FracDefi}
\end{equation}%
It is built as a nonlocal operator, produced by the juxtaposition of the
direct and inverse Fourier transforms, with the fractional differentiation
represented by factor $|p|^{\alpha }$ in the Fourier space. In the Laskin's
fractional quantum mechanics, $\alpha $ takes values
\begin{equation}
1<\alpha \leq 2.  \label{interval}
\end{equation}%
The limit of $\alpha =2$ corresponds to the canonical Schr\"{o}dinger
equation, with operator (\ref{FracDefi}) reducing\ to the normal second
derivative. The case of $\alpha =1$ (the \textquotedblleft square root of
the Laplacian") is known as the Cauchy driver \cite{Cauchy,Cauchy2}, which
occurs, in particular, in the fractional equation for a photonic wave
function \cite{photonic}. Further, it was found that the original LI\ range (%
\ref{interval}) may be expanded to $0<\alpha \leq 2$ \cite{photonic}.

It is relevant to mention that there are other definitions of the
fractional-order differentiation, such as the Caputo derivative \cite%
{Baleanu}, but the derivation of the fractional Schr\"{o}dinger equation by
Laskin yields the model with the Riesz derivative \cite{Lask1,Lask2}. The
nonlocal character of the fractional-derivative operators produces
essentially nontrivial consequences in bounded domains, with the respective
boundary conditions \cite{Cauchy2,Garbaczewski-PRE}. In particular,
different definitions of the fractional derivative produce essentially
non-equivalent results in the case of boundary conditions and under the
action of external potentials.

While experimental realization of the fractional quantum mechanics is still
missing, it was proposed to emulate it by means of the classical wave
propagation in photonics, utilizing the commonly known similarity of the Schr%
\"{o}dinger equation and the equation for the paraxial propagation of
optical beams \cite{EXP3,PROP}. The theoretically elaborated setup for the
emulation of the fractional diffraction is based on the 4\textit{f} optical
configuration, which performs the spatial Fourier decomposition of the beam
by means of a lens, then carries the beam through an appropriately shaped
phase plate to impart the local phase shift which emulates the action of the
fractional diffraction as per Eq. (\ref{FracDefi}), and finally retrieves
the beam from its Fourier decomposition, using another lens \cite{EXP3}.
Other emulations of the fractional quantum mechanics were proposed in L\'{e}%
vy crystals \cite{EXP1} and polariton condensates \cite{EXP2}.

Very recently, the first \emph{experimental realization} of the effective
fractional group-velocity dispersion (rather than transverse diffraction) in
a fiber-laser setup has been reported \cite{Shilong}. While the fractional
dispersion acts in the temporal domain, its experimental emulation was
carried out by means of a 4\textit{f} setup similar to the one outlined
above, with the difference that the lenses performed the spectral
decomposition of the optical signal and inverse recombination. To avoid
misunderstanding, it is relevant to stress that the temporal variable plays
the role of an efficient coordinate in optical fibers, while the evolution
variable is the propagation distance. Therefore, the effective fractional
dispersion reported in Ref. \cite{Shilong} emulates the temporal counterpart
of the fractional diffraction (cf. Ref. \cite{Jorge}), but not a fractional
derivative with respect to the evolution variable, which occurs in various
models of anomalous diffusion \cite{time1,time2}. Of course, a caveat is
that the proposed and reported realizations of the fractional diffraction
and dispersion in optics actually report the simulation of these effects by
light, but not their direct action.

Theoretical studies of models with the fractional diffraction were developed
by including external potentials [in particular, parity-time ($\mathcal{PT}$%
) symmetric ones \cite{PTS}-\cite{PT coupler}] and studying the propagation
of Airy waves in the two-dimensional geometry, with the
fractional-diffraction operator $\left( -\partial ^{2}/\partial
x^{2}-\partial ^{2}/\partial y^{2}\right) ^{\alpha /2}$ \cite%
{Yingji1,Yingji2}. The implementation of the fractional
diffraction/dispersion in optical waveguides makes it natural to include the
self-focusing Kerr nonlinearity of the material, which leads to the
consideration of fractional nonlinear Schr\"{o}dinger equations (FNLSEs)
\cite{review}.

The work based on FNLSEs has produced many theoretical results, including
the modulational instability of continuous waves \cite{Conti}, critical or
supercritical collapse \cite{Chen}, and diverse species of solitons \cite%
{Jorge1}-\cite{review}. These are quasi-linear \textquotedblleft accessible
solitons" \cite{Frac1,Frac2}, gap solitons maintained by lattice potentials
\cite{Frac5a}-\cite{Frac5}, solitary vortices \cite{Frac6,Frac7,HS},
multi-peak modes \cite{Frac8}-\cite{Frac11}, clusters \cite{Frac12},
discrete solitons \cite{Frac13}, and dark modes \cite{we}. Solitons produced
by the interplay of the fractional diffraction and quadratic nonlinearity
were predicted too \cite{Thirouin,quadratic}, as well as dissipative
solitons produced by the fractional complex Ginzburg-Landau equation \cite%
{Frac14}.

A generic effect produced by self-trapping nonlinearity is spontaneous
symmetry breaking (SSB) in double-well potentials \cite{book}. SSB
phenomenology in such settings was studied in detail theoretically \cite%
{Davies}-\cite{Warsaw} and experimentally \cite{lasers}-\cite{nano} in
diverse physical settings. Recently, the theoretical analysis was extended
for nonlinear systems combining the fractional diffraction and symmetric
potentials \cite{Frac16}-\cite{soliton5}. An appropriate platform for the
realization of SSB in the form of two-component solitons with broken
symmetry between the components is offered by dual waveguides, which are
often represented by double-core optical fibers \cite{progress,Peng}. The
transition from symmetric to asymmetric two-component solitons, i.e., the
\textit{SSB bifurcation} in such fibers, was studied in detail theoretically
\cite{Wabnitz}-\cite{Pak}, and it was recently demonstrated in an experiment
\cite{Bugar}. In another recent work \cite{Zeng-Zeng}, some families of
symmetric and asymmetric solitons were found in the double-core system with
fractional diffraction.

The objective of the present work is to identify the SSB bifurcation of
two-component solitons in the fractional dual-core waveguide, i.e., the
corresponding symmetry-breaking phase transition. We do this analytically,
by means of the variational approximation (VA), and in a systematic
numerical form. An essential finding is that \emph{deeper} system's
fractionality [i.e., smaller LI $\alpha $ in Eq. (\ref{FracDefi})] enhances
the \emph{subcritical} character \cite{bif} of the bifurcation. In other
words, the fractionality makes the SSB of the two-component solitons a more
strongly pronounced phase transition of the first kind.

An essential peculiarity of the fractional diffraction is that it does not
admit the Galilean invariance. Therefore, the generation of moving solitons
and the consideration of the SSB bifurcation for them is a nontrivial
problem, which we address here too, and which was not considered previously.
Collisions between moving solitons are studied too

Thus, we address the system of linearly-coupled FNLSEs with the cubic
self-focusing nonlinearity,%
\begin{eqnarray}
i\frac{\partial u_{1}}{\partial t} &=&\frac{1}{2}\left( -\frac{\partial ^{2}%
}{\partial x^{2}}\right) ^{\alpha /2}u_{1}-|u_{1}|^{2}u_{1}-u_{2},  \notag \\
i\frac{\partial u_{2}}{\partial t} &=&\frac{1}{2}\left( -\frac{\partial ^{2}%
}{\partial x^{2}}\right) ^{\alpha /2}u_{2}-|u_{2}|^{2}u_{2}-u_{1},
\label{1D}
\end{eqnarray}%
where the coupling coefficient in front of terms $-u_{2}$ and $-u_{1}$ is
fixed to be $1$ by means of scaling. The only irreducible control parameter
of the system is LI $\alpha $. In terms of optics, Eqs. (\ref{1D}) describe
a pair of parallel planar waveguides coupled by tunneling of light. In this
case, $t$ and $x$ are, respectively, the propagation distance and transverse
coordinate. The same system realized in the temporal domain, i.e., with $x$
standing for the reduced time \cite{Agrawal}, may be construed as a model of
a laser built of dual-core fibers with the effective fractional dispersion,
following the experimental method reported in Ref. \cite{Shilong}. The
possibility of using dual-core fibers in lasers is a well-known option \cite%
{Winful}.

There may also be a possibility to consider Eqs. (\ref{1D}) as a system of
coupled Gross-Pitaevskii equations for a Bose-Einstein condensate of quantum
particles governed by the fractional Schr\"{o}dinger equations and filling a
pair of parallel cigar-shaped traps, which are linearly coupled by tunneling
of the particles. In that case, $t$ is time and $x$ is the coordinate along
the traps. However, the derivation of such a mean-field model for the
condensate requires accurate analysis, which should be a subject of a
separate work. In particular, it is necessary to check if collisional
effects in this system may be properly represented by the usual mean-field
cubic terms. Experimental realization of the system may be a challenging
objective too.

The presentation is organized below as follows. The framework for the
construction of soliton solutions and analysis of their stability and
dynamics are presented in Section 2. The analytical approach, based on the
VA, is developed in Section 3. Numerical results for quiescent and moving
solitons are summarized in Section 4. The work is concluded by Section 5.

\section{The framework for soliton solutions}

Stationary-state solutions to Eq. (\ref{1D}) with propagation constant $k$
(in the application to BEC, $-k$ is the chemical potential) are looked for as%
\begin{equation}
u_{1,2}(x,t)=U_{1,2}(x)e^{ikt},  \label{U12}
\end{equation}%
with real functions $U_{1,2}(x)$ satisfying equations%
\begin{eqnarray}
kU_{1}+\frac{1}{2}\left( -\frac{\partial ^{2}}{\partial x^{2}}\right)
^{\alpha /2}U_{1}-U_{1}^{3}-U_{2} &=&0,  \notag \\
kU_{2}+\frac{1}{2}\left( -\frac{\partial ^{2}}{\partial x^{2}}\right)
^{\alpha /2}U_{2}-U_{2}^{3}-U_{1} &=&0.  \label{UU}
\end{eqnarray}%
Taking into regard that $U_{1,2}(x)$ are even functions of $x$ and using the
definition of the fractional derivative given by Eq. (\ref{FracDefi}), the
explicit form of Eq. (\ref{UU}) can be written as%
\begin{eqnarray}
kU_{1}+\frac{1}{2\pi }\int_{0}^{+\infty }p^{\alpha }dp\int_{-\infty
}^{+\infty }\cos (px)\cos (px^{\prime })U_{1}(x^{\prime })dx^{\prime
}-U_{1}^{3}-U_{2} &=&0,  \notag \\
kU_{2}+\frac{1}{2\pi }\int_{0}^{+\infty }p^{\alpha }dp\int_{-\infty
}^{+\infty }\cos (px)\cos (px^{\prime })U_{2}(x^{\prime })dx^{\prime
}-U_{2}^{3}-U_{1} &=&0.  \label{UUsimple}
\end{eqnarray}%
We consider values of LI in the interval of
\begin{equation}
1<\alpha \leq 2,  \label{1-2}
\end{equation}%
as it is well known that, at $\alpha \leq 1$, the FNLSE gives rise to the
collapse (critical collapse at $\alpha =1$, and supercritical at $\alpha <1$%
) \cite{Chen,review}.

In the case of the normal diffraction, $\alpha =2$, obvious solutions of Eq.
(\ref{UU}) in the form of symmetric solitons is%
\begin{equation}
U_{1}=U_{2}=\sqrt{2(k-1)}\mathrm{sech}\left( \sqrt{2(k-1)}x\right) .
\label{symm}
\end{equation}%
The norm (power) of this solution is%
\begin{equation}
N=\int_{-\infty }^{+\infty }\left[ \left( U_{1}(x)\right) ^{2}+\left(
U_{2}(x)\right) ^{2}\right] dx=4\sqrt{2\left( k-1\right) }.  \label{N}
\end{equation}%
With the increase of $N$, the symmetric states become unstable through SSB,
and stable asymmetric solitons appear. While there are no exact solutions
for the asymmetric solitons, the SSB point, at which they emerge, can be
found exactly for $\alpha =2$ \cite{Wabnitz}:%
\begin{equation}
\left( N_{\mathrm{SSB}}\right) _{\mathrm{exact}}(\alpha =2)=8/\sqrt{3}.
\label{SSB-exact}
\end{equation}%
On the other hand, the VA predicts this point at%
\begin{equation}
\left( N_{\mathrm{SSB}}\right) _{\mathrm{VA}}(\alpha =2)=2\sqrt{6}\approx
4.899  \label{SSB-VA}
\end{equation}%
\cite{Pak,progress}, the relative error of this result being $\simeq 6\%$.

Asymmetry of the solitons produced by the SSB bifurcation is defined by
parameter%
\begin{equation}
\Theta =N^{-1}\int_{-\infty }^{+\infty }\left[ \left( U_{1}(x)\right)
^{2}-\left( U_{2}(x)\right) ^{2}\right] dx.  \label{Theta}
\end{equation}%
The bifurcation is characterized by diagrams which displays $\Theta $ as
functions of $k$ or $N$ [see, in particular, Fig. \ref{theta} below].

Solutions of Eq. (\ref{1D}) for moving solitons, i.e., obliquely propagating
light beams with slope $c$ in the planar waveguide, are sought for as
\begin{equation}
u_{1,2}=u_{1,2}\left( \xi \equiv x-ct,t\right) .  \label{xi0}
\end{equation}%
Accordingly, Eq. (\ref{1D}) is rewritten in terms of $\left( \xi ,t\right) $
as
\begin{eqnarray}
i\frac{\partial u_{1}}{\partial t}-ic\frac{\partial u_{1}}{\partial \xi } &=&%
\frac{1}{2}\left( -\frac{\partial ^{2}}{\partial \xi ^{2}}\right) ^{\alpha
/2}u_{1}-|u_{1}|^{2}u_{1}-u_{2},  \notag \\
i\frac{\partial u_{2}}{\partial t}-ic\frac{\partial u_{2}}{\partial \xi } &=&%
\frac{1}{2}\left( -\frac{\partial ^{2}}{\partial \xi ^{2}}\right) ^{\alpha
/2}u_{2}-|u_{2}|^{2}u_{2}-u_{1}.  \label{cc}
\end{eqnarray}%
Solutions to Eq. (\ref{cc}) are further looked for as $u_{1,2}\left( \xi
,t\right) =$ $U_{1,2}(\xi )e^{ikt}$ [cf. Eq. (\ref{U12})], with complex
functions $U_{1,2}(\xi )$ satisfying the following system of stationary
equations:%
\begin{eqnarray}
kU_{1}+ic\frac{dU_{1}}{d\xi }+\frac{1}{2}\left( -\frac{d^{2}}{d\xi ^{2}}%
\right) ^{\alpha /2}U_{1}-\left\vert U_{1}\right\vert ^{2}U_{1}-U_{2} &=&0,
\notag \\
kU_{2}+ic\frac{dU_{2}}{d\xi }+\frac{1}{2}\left( -\frac{d^{2}}{d\xi ^{2}}%
\right) ^{\alpha /2}U_{2}-\left\vert U_{2}\right\vert ^{2}U_{2}-U_{1} &=&0.
\label{cc-stationary}
\end{eqnarray}

The stability of solitons was addressed by considering solutions including
small perturbations $a_{1,2}$ and $b_{1,2}$,
\begin{equation}
u_{1,2}(x,t)=\left[ U_{1,2}(x)+a_{1,2}(x)e^{\lambda t}+b_{1,2}^{\ast
}(x)e^{\lambda ^{\ast }t}\right] e^{ikt}\,,  \label{pert}
\end{equation}%
where $\lambda $ is the instability growth rate, and $\ast $ stands for the
complex conjugate. The linearization of Eq. (\ref{1D}) for the perturbations
leads the system of the corresponding Bogoliubov -- de Gennes equations:%
\begin{eqnarray}
\left[ -\left( k-i\lambda \right) -\frac{1}{2}\left( -\frac{d^{2}}{d\xi ^{2}}%
\right) ^{\alpha /2}+2\left\vert U_{1,2}\right\vert ^{2}\right]
a_{1,2}+U_{1,2}^{2}b_{1,2}+a_{2,1} &=&0,  \notag \\
\left[ -\left( k+i\lambda \right) -\frac{1}{2}\left( -\frac{d^{2}}{d\xi ^{2}}%
\right) ^{\alpha /2}+2\left\vert U_{1,2}\right\vert ^{2}\right]
b_{1,2}+\left( U_{1,2}^{\ast }\right) ^{2}a_{1,2}+b_{2,1} &=&0.  \label{BdG}
\end{eqnarray}%
The stability condition is that solutions of Eq. (\ref{BdG}) must produce
only eigenvalues with Re$\left( \lambda \right) =0$ \cite{W,Yang}. Below,
the prediction for the stability of the solitons, provided by these
equations, is corroborated by direct simulations of the evolution of
perturbed solitons.

\section{The variational approximation (VA)}

To apply VA, we note that Eq. (\ref{UUsimple}) can be derived from the
Lagrangian,%
\begin{equation}
L=\int_{-\infty }^{+\infty }\left[ \frac{k}{2}\left(
U_{1}^{2}+U_{2}^{2}\right) \right] dx+H,  \label{L}
\end{equation}%
with Hamiltonian%
\begin{gather}
H=\int_{-\infty }^{+\infty }\left\{ -\frac{1}{4}\left[
U_{1}^{4}(x)+U_{2}^{4}(x)\right] -U_{1}(x)U_{2}(x)\right\} dx  \notag \\
+\frac{1}{4\pi }\int_{0}^{+\infty }p^{\alpha }dp\int_{-\infty }^{+\infty
}dx\int_{-\infty }^{+\infty }dx^{\prime }\cos \left( p\left( x-x^{\prime
}\right) \right) \left[ U_{1}(x)U_{1}(x^{\prime })+U_{2}(x)U_{2}(x^{\prime })%
\right] ,  \label{H}
\end{gather}%
cf. Ref. \cite{HS}. The ansatz for the asymmetric soliton can be adopted in
the simple form, which follows the pattern of the above-mentioned solution (%
\ref{symm}):%
\begin{equation}
U_{1}(x)=\sqrt{\frac{N}{2W}}\left( \cos \chi \right) \mathrm{sech}\left(
\frac{x}{W}\right) ,U_{2}(x)=\sqrt{\frac{N}{2W}}\left( \sin \chi \right)
\mathrm{sech}\left( \frac{x}{W}\right) ,  \label{ans}
\end{equation}%
where variational parameters are width $W$ and norm-distribution angle $\chi
$, while $N$ is considered as a given total norm. The asymmetry parameter (%
\ref{Theta}) corresponding to the ansatz is%
\begin{eqnarray}
\Theta _{\mathrm{VA}} &=&\cos \left( 2\chi \right) \equiv \sqrt{1-S^{2}},
\label{ThetaVA} \\
S &\equiv &\sin \left( 2\chi \right)  \label{Ssin}
\end{eqnarray}%
(parameter $S$ is used below). The substitution of ansatz (\ref{ans}) in
Lagrangian defined by Eqs. (\ref{L}) and (\ref{H}) yields%
\begin{eqnarray}
L_{\mathrm{VA}} &=&\frac{N}{2}k-\frac{N^{2}}{12W}\left( 1-\frac{1}{2}\sin
^{2}(2\chi )\right) -\frac{N}{2}\sin (2\chi )  \notag \\
&&+\left( 1-2^{1-\alpha }\right) \Gamma (1+\alpha )\zeta (\alpha )\frac{N}{%
2(\pi W)^{\alpha }},  \label{Leff}
\end{eqnarray}%
where $\Gamma $ and $\zeta $ are the Gamma- and zeta-functions. Parameters
of the asymmetric solitons are predicted by the Euler-Lagrange equations,%
\begin{equation}
\frac{\partial L_{\mathrm{VA}}}{\partial W}=\frac{\partial L_{\mathrm{VA}}}{%
\partial \left( \sin (2\chi \right) )}=0,  \label{EL}
\end{equation}%
which amount to relation $W=(N/6)\sin (2\chi )$, and an equation for $%
S\equiv \sin \left( 2\chi \right) $:%
\begin{equation}
S^{\alpha -1}\left( 1-\frac{S^{2}}{2}\right) =\frac{\alpha }{\pi ^{\alpha }}%
\left( 1-2^{1-\alpha }\right) \Gamma (1+\alpha )\zeta (\alpha )\left( \frac{6%
}{N}\right) ^{\alpha }.  \label{S}
\end{equation}

In particular, the threshold at which SSB takes place, giving rise to
families of asymmetric solutions which branch off from symmetric ones, that
correspond to $S=1$, is produced by the substitution of $S=1$ in Eq. (\ref{S}%
):%
\begin{equation}
\left( N_{\mathrm{SSB}}\right) _{\mathrm{VA}}(\alpha )=\frac{6}{\pi }\left[
2\alpha \left( 1-2^{1-\alpha }\right) \Gamma (1+\alpha )\zeta (\alpha )%
\right] ^{1/\alpha }.  \label{N_SSB}
\end{equation}%
In the case of $\alpha =2$, expression (\ref{N_SSB}) exactly reduces to Eq. (%
\ref{SSB-VA}). In the opposite limit of $\alpha -1\rightarrow +0$ [see Eq. (%
\ref{1-2})], Eq. (\ref{N_SSB}) yields%
\begin{equation}
\left( N_{\mathrm{SSB}}\right) _{\mathrm{VA}}\left( \alpha \rightarrow
1\right) =12\pi ^{-1}\ln 2\approx 2.648.  \label{alpha=1}
\end{equation}%
%

Finally, the VA prediction for the asymmetry dependence on the norm, $\Theta
(N)$, is obtained, in an implicit form, as a combination of Eqs. (\ref%
{ThetaVA}), (\ref{Ssin}), and (\ref{S}):%
\begin{equation}
\left( 1-\Theta _{\mathrm{VA}}^{2}\right) ^{(\alpha -1)/2}\left( 1+\Theta _{%
\mathrm{VA}}^{2}\right) =\frac{2\alpha }{\pi ^{\alpha }}\left( 1-2^{1-\alpha
}\right) \Gamma (1+\alpha )\zeta (\alpha )\left( \frac{6}{N}\right) ^{\alpha
}.  \label{Th(N)}
\end{equation}%
This relation takes an explicit form in the limit of $\alpha \rightarrow 1$:%
\begin{equation}
\Theta _{\mathrm{VA}}(N;\alpha \rightarrow 1)=\sqrt{\frac{\left( N_{\mathrm{%
SSB}}\right) _{\mathrm{VA}}\left( \alpha \rightarrow 1\right) }{N}-1},
\label{extreme}
\end{equation}%
where $\left( N_{\mathrm{SSB}}\right) _{\mathrm{VA}}\left( \alpha
\rightarrow 1\right) $ is the value given by Eq. (\ref{alpha=1}). Dependence
(\ref{extreme}) is displayed below in Fig. \ref{theta}, and the predictions
produced by Eqs. (\ref{Th(N)}) and (\ref{N_SSB}) are compared with numerical
results in Fig. \ref{theta}.

\section{Numerical results}

\subsection{The spontaneous-symmetry-breaking (SSB) bifurcation of
stationary states and their stability}

\begin{figure}[tbph]
\subfigure[]{%
\includegraphics[scale=0.4]{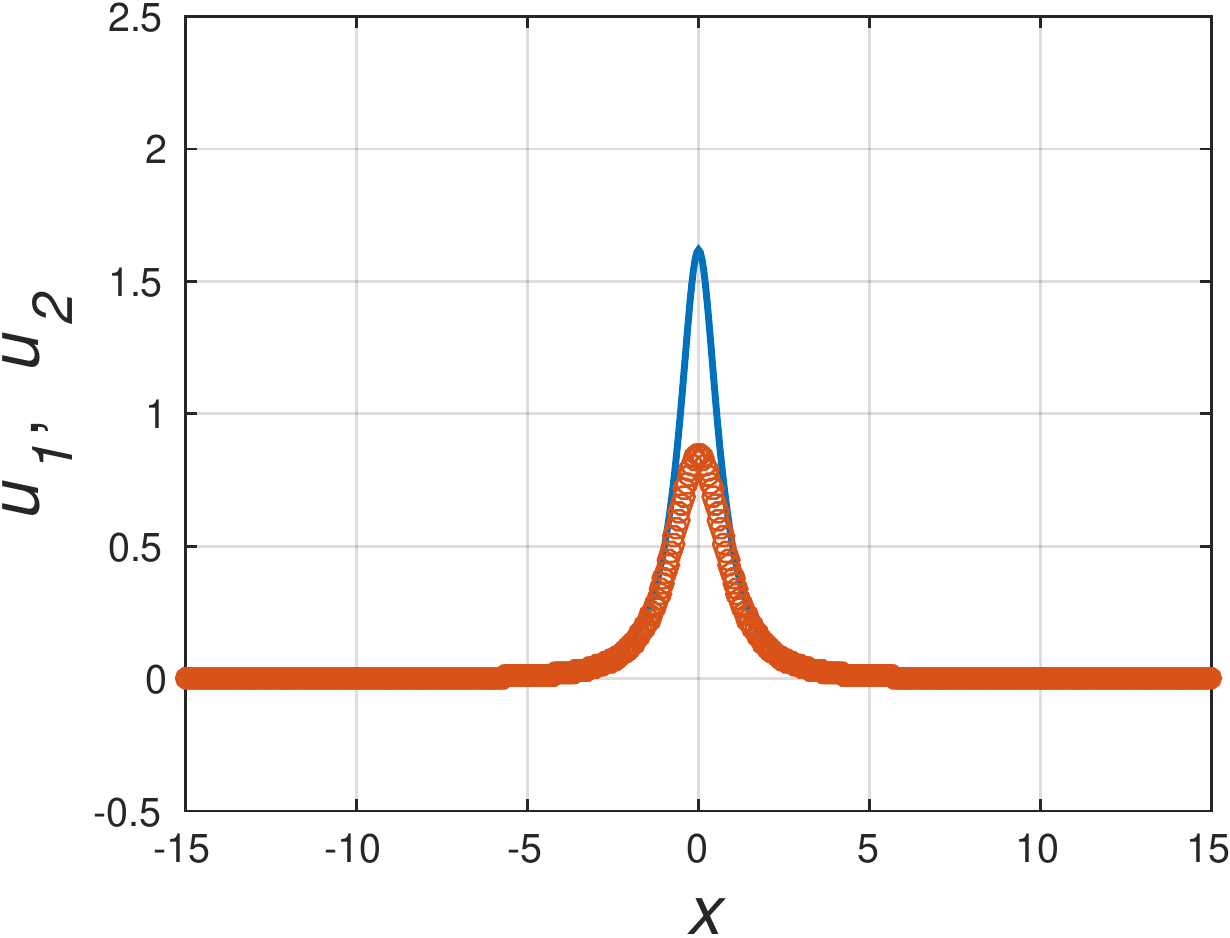}} %
\subfigure[]{%
\includegraphics[scale=0.4]{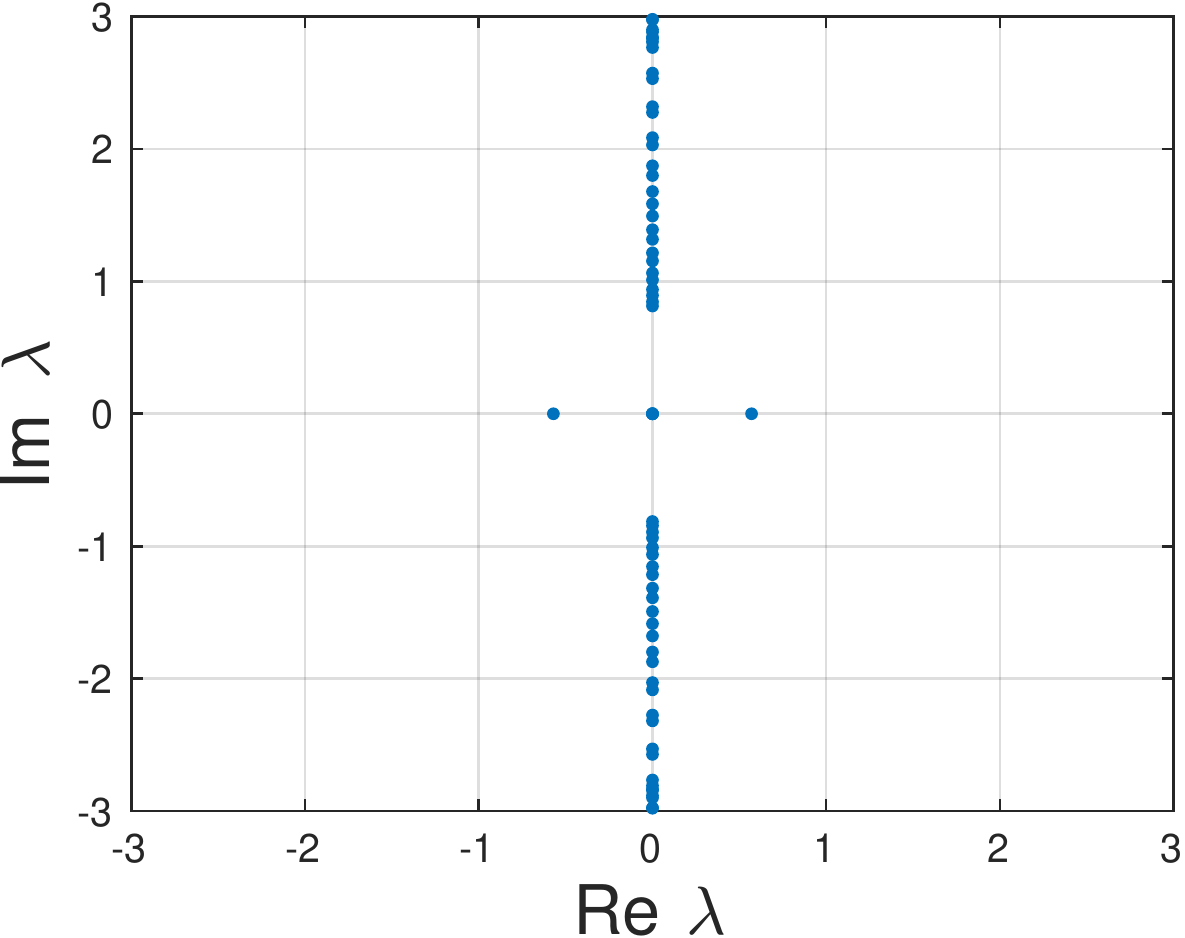}} %
\subfigure[]{%
\includegraphics[scale=0.5]{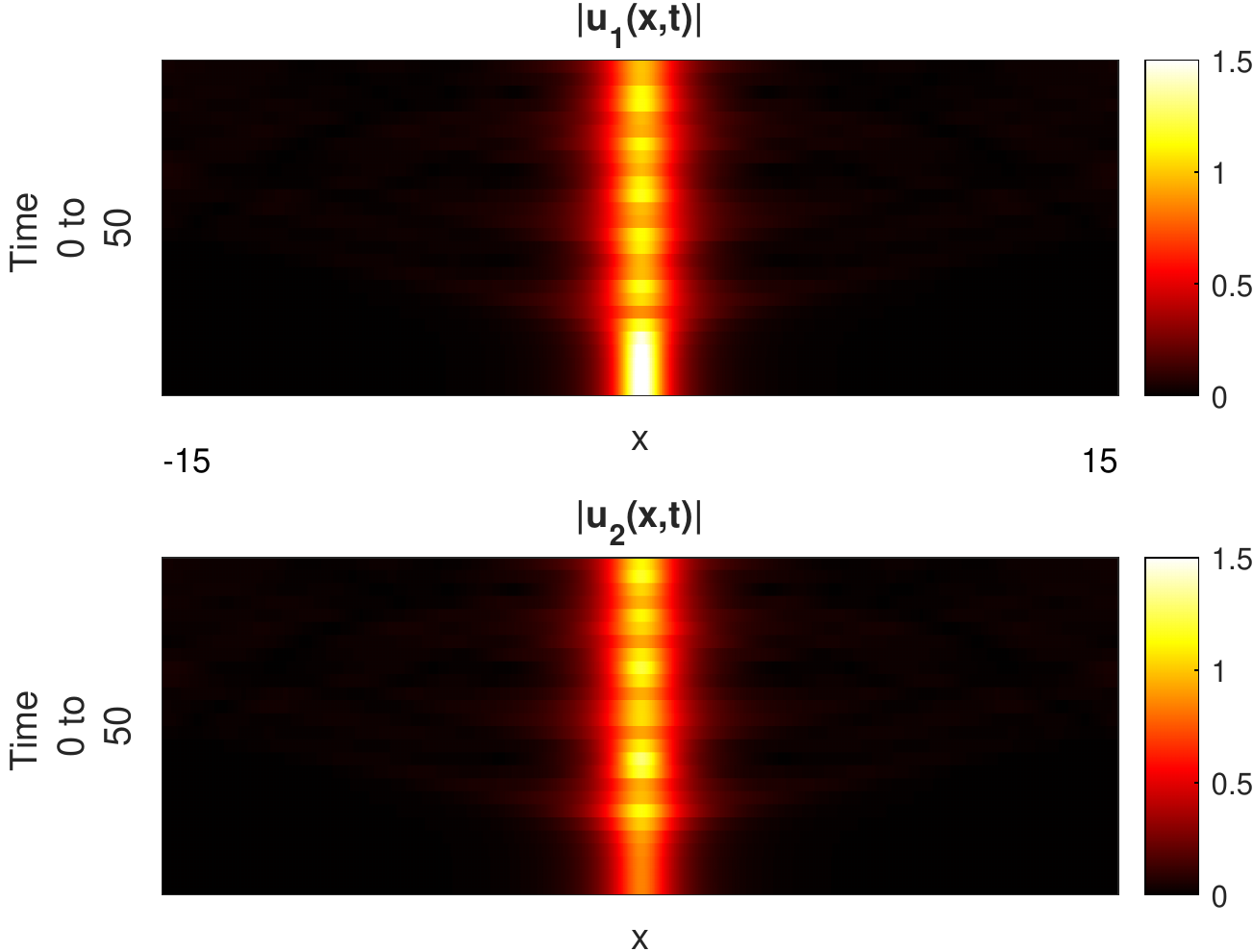}}
\caption{The stationary profile of an unstable asymmetric soliton (a), its
spectrum of perturbation eigenvalues (b), and perturbed evolution (c), for $%
\protect\alpha =1.6$, $k=1.8$, and $N=3.729$. Shapes of stable asymmetric
solitons are similar to the one displayed here.}
\label{asym-unstable-map}
\end{figure}

Soliton solutions of Eq. (\ref{UUsimple}) were produced by means of the
squared-operator iteration method \cite{YangLakoba,Yang}. Then, the spectrum
of stability eigenvalues $\lambda $ was produced solving Eq. (\ref{BdG}) by
means of the Fourier collocation method. Both algorithms were realized in
the Matlab shell, as outlined in~Ref. \cite{Yang}. Direct simulations of Eq.
(\ref{1D}) were performed by means of the pseudospectral method~\cite%
{Yang,Trefethen,Boyd,Guo},

A typical profile of an asymmetric soliton is presented in Fig. \ref%
{asym-unstable-map}(a). Panels (b) and (c) of the figure demonstrate that
this soliton is unstable, spontaneously transforming into a robust breather,
which is a dynamical state effectively symmetrized by persistent
oscillations between its two components. The shape of stable asymmetric
soliton is similar to that in Fig. \ref{asym-unstable-map}. As for symmetric
solitons, those ones which are unstable spontaneously turn into stable
asymmetric ones, with residual internal oscillations (not shown here in
detail). 

\begin{figure}[tbph]
\subfigure[]{\includegraphics[scale=0.5]{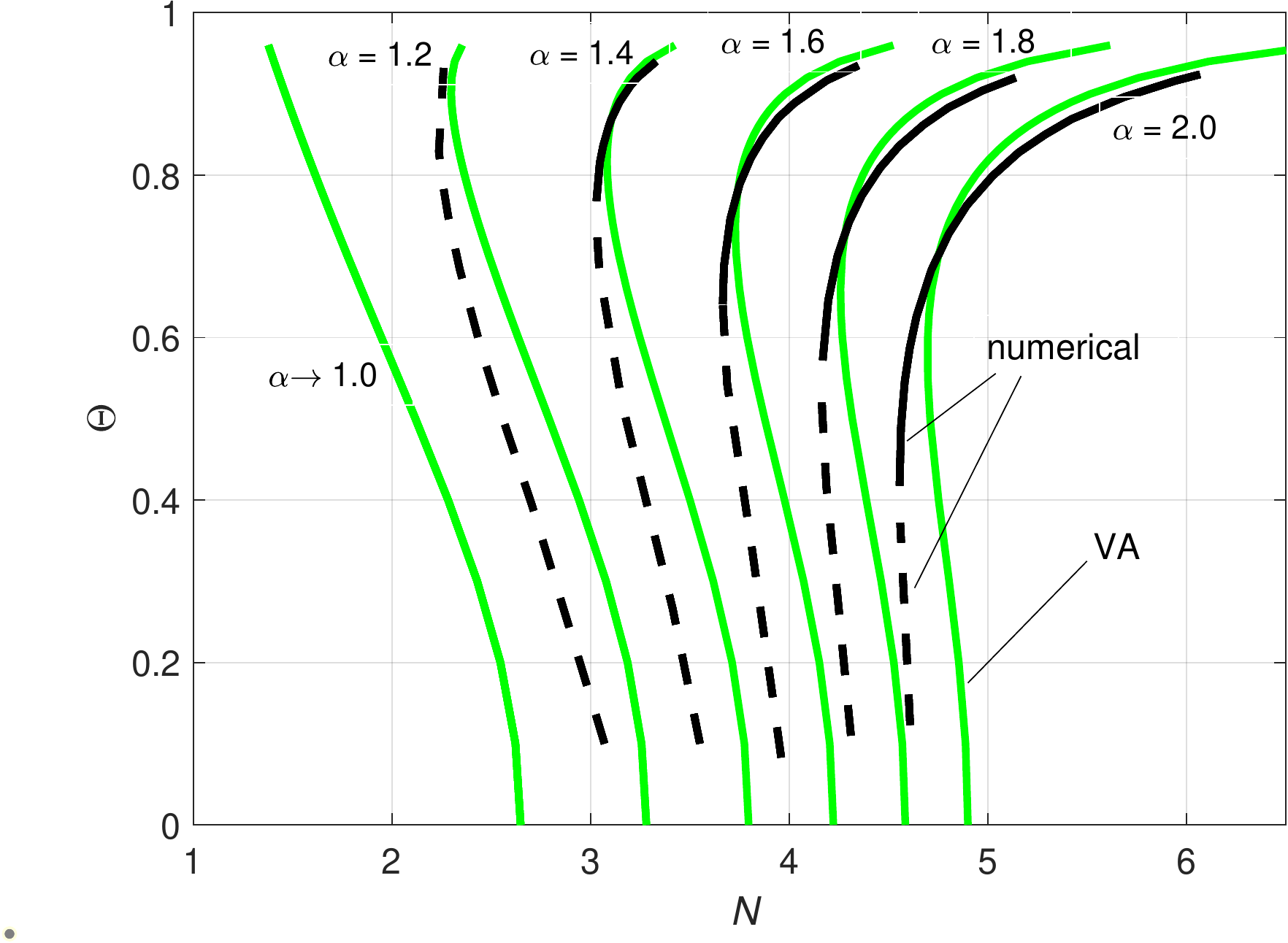}} \subfigure[]{%
\includegraphics[scale=0.5]{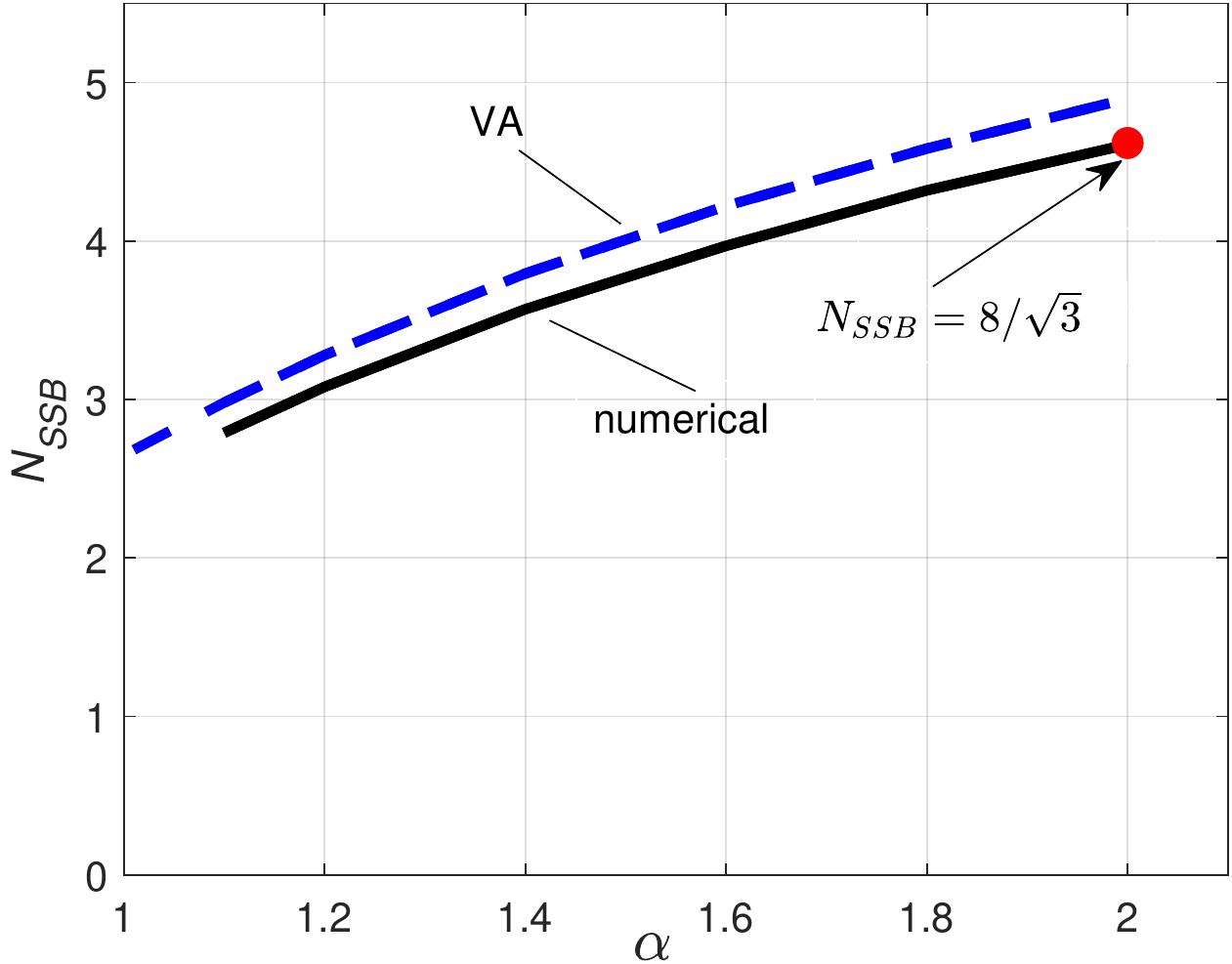}} 
\caption{(a) Symmetry parameter $\Theta $, defined as per Eq. (\protect\ref%
{Theta}), versus the norm, for families of asymmetric solitons, at indicated
values of LI $\protect\alpha $. Plotted are both the numerical results and
their analytical counterparts, predicted by VA as per Eq. (\protect\ref%
{Th(N)}). For $\protect\alpha \rightarrow 1$, the VA curve is plotted
according to Eq. (\protect\ref{extreme}). Solid and dashed segments of the
numerically generated branches mark stable and unstable solutions,
respectively. (b) The value of the norm at the SSB bifurcation point versus
LI $\protect\alpha $, as produced by the numerical solution and predicted by
VA, see Eq. (\protect\ref{N_SSB}). At $\protect\alpha =2$, the red point
shows the exact value (\protect\ref{SSB-exact}), which is identical to the
corresponding numerically found one. At $\protect\alpha =1$, the VA value is
given by Eq. (\protect\ref{alpha=1}).}
\label{theta}
\end{figure}

Families of numerically found stationary symmetric and asymmetric solitons,
and the SSB bifurcation which couples them, are displayed in Figs. \ref%
{bifur} and \ref{theta} by respective dependences $N(k)$ and $\Theta (N)$,
in the interval of LI values $1.1\leq \alpha \leq 2.0$, cf. Eq. (\ref{1-2})
(the numerical solution is technically challenging for $\alpha <1.1$ because
of slow convergence). The latter figure includes the comparison with the VA
prediction, given above by Eq. (\ref{Th(N)}). It is seen that the VA is
reasonably accurate, with the relative discrepancy in terms of $N$ for fixed
$\Theta $ being $\leq 6\%$ in Fig. \ref{theta}(a). In particular, the VA
results are quite reliable for stable branches $\Theta (N)$. In the case of $%
\alpha =2$, the findings are tantamount to the well-known results for the
usual coupler \cite{progress,Peng}.

Note that the (in)stability of all solution branches, as shown in Figs. \ref%
{bifur} and \ref{theta}, complies with the Vakhitov-Kolokolov criterion, $%
dN/dk>0$, which is the necessary stability condition for self-trapped modes
\cite{VK,Yang}. In particular, the asymmetric solitons belonging to the
backward- and forward-going segments of the respective solution branches are
stable or not in agreement with the criterion.

\begin{figure}[tbph]
\begin{center}
\includegraphics[scale=0.5]{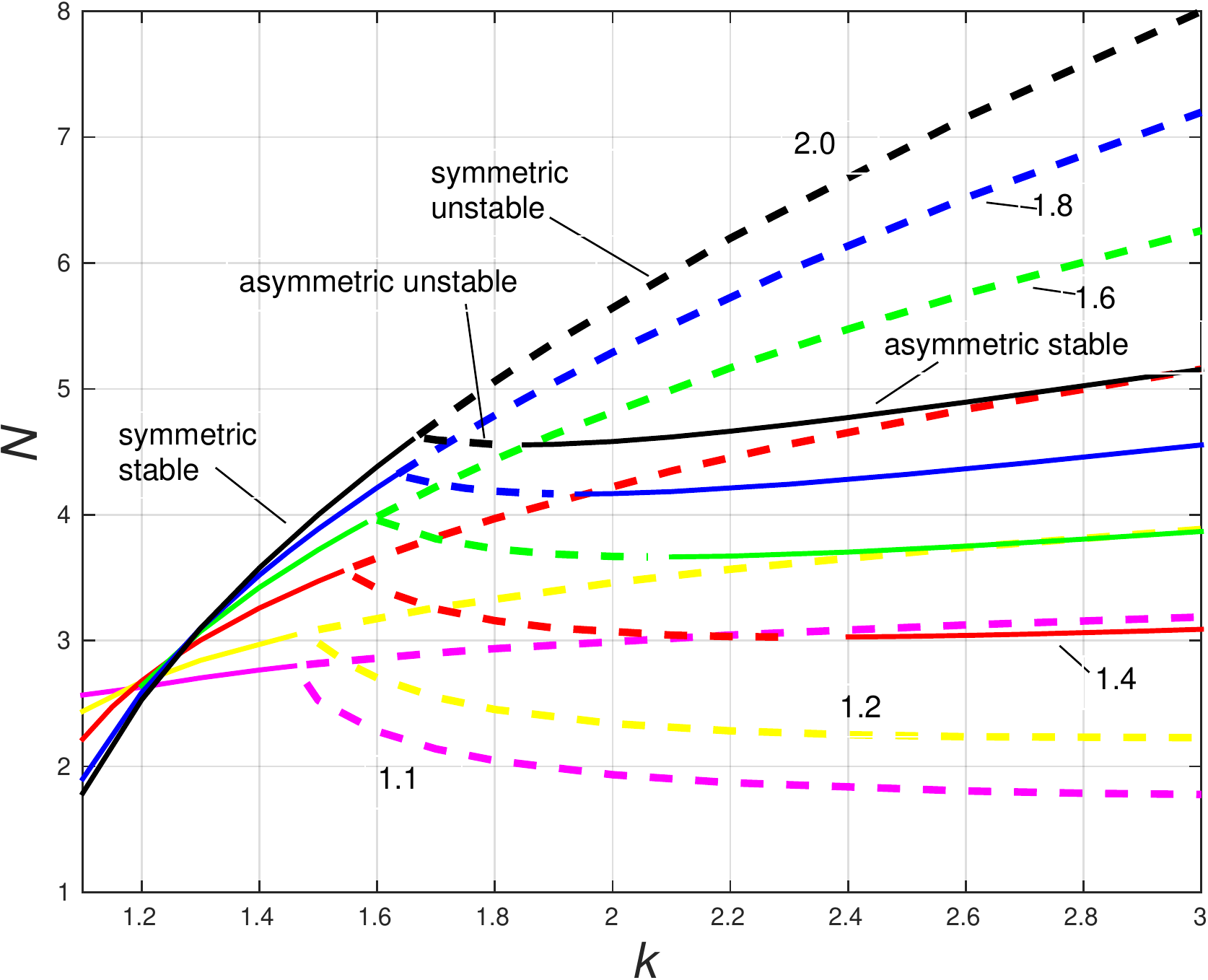}
\end{center}
\caption{Norm $N$ of symmetric and asymmetric solitons plotted versus the
propagation constant, $k$, at different values of LI $\protect\alpha $, as
indicated by labels. Solid and dashed lines mark stable and unstable
branches, respectively, according to the eigenvalue spectrum produced by Eq.
(\protect\ref{BdG}).}
\label{bifur}
\end{figure}

Figures \ref{bifur} and \ref{theta}(a) demonstrate that, as it might be
expected, the increase of the norm leads to destabilization of symmetric
solitons and emergence of asymmetric ones via the SSB\ bifurcation. Similar
to the known feature of SSB in the usual coupler ($\alpha =2$), Fig. \ref%
{theta} shows that the bifurcation is of the subcritical type \cite{bif},
thus representing a phase transition of the first kind, which admits
hysteresis and bistability. The subcritical bifurcation gives rise to
branches of asymmetric solitons that originally go in the backward direction
(which corresponds to the decrease of $N$), and then turn forward at
critical points. These points represent the minimum value of $N$ at which
the asymmetric solitons exist. An essential conclusion suggested by Fig. \ref%
{theta} is that \emph{deeper fractionality}, i.e., larger $2-\alpha $, makes
the subcritical character of the SSB bifurcation stronger, while this
feature is very weak in the usual case, $\alpha =2$. As an extension of this
trend, Fig. \ref{theta} shows that, at $\alpha =1.2$, the branch of the
asymmetric solitons admits almost no extension past the critical turning
point. Additional numerical results demonstrate that, in the explored range
of values of $k$, the $\Theta (N)$ curve does not reach the turning point
for $\alpha =1.1$. A conjecture is that the SSB bifurcation becomes an \emph{%
extreme subcritical one}, with no turning points, in the limit of $\alpha
\rightarrow 1$. Indeed, this feature is explicitly demonstrated by the VA
curve produced by Eq. (\ref{extreme}), which is plotted in Fig. \ref{theta}%
(a).

Note that the first example of the extreme subcritical SSB bifurcation was
reported in the model based on the single NLS equation with the usual
diffraction and nonlinear double-well potential \cite{Dong},%
\begin{equation}
i\frac{\partial u}{\partial t}=-\frac{1}{2}\frac{\partial ^{2}u}{\partial
x^{2}}-\left[ \delta \left( x-1\right) +\delta \left( x+1\right) \right]
|u|^{2}u,  \label{delta}
\end{equation}%
where $\delta $ is the delta-function, with separation $2$ between the
potential wells fixed by scaling. An exact analytical solution of Eq. (\ref%
{delta}) produces the extreme subcritical bifurcation, with branches of
asymmetric solitons going backward from the bifurcation point, $N(\Theta =0)=%
\frac{2}{3}+\frac{8}{27}\left( \frac{3}{4}+\ln 2\right) \approx 1.09$, up to
$N(\Theta =1)=1$, and never turning forward.

The basic characteristic of the SSB is the value of the norm, $N_{\mathrm{SSB%
}}$, at the bifurcation point, which is predicted by VA as per Eq. (\ref%
{N_SSB}).
The numerically found value is displayed, as a function of $\alpha $, along
with its VA-predicted counterpart, in Fig. \ref{theta}(b). The mismatch
between the numerical and variational values does not exceed $6\%$.

In addition to the symmetric and asymmetric solitons, solutions for
antisymmetric ones, with $U_{1}(x)=-U_{2}(x)$, were constructed too, but
they are completely unstable (similar to the known situation in the usual
coupler, with $\alpha =2$ \cite{progress,Peng}). This conclusion is
explained by the fact that the linear-coupling term in Hamiltonian (\ref{H}%
), with density $-U_{1}(x)U_{2}(x)$, is negative for symmetric and
asymmetric modes, but positive for the antisymmetric ones, the higher energy
(Hamiltonian) implying the instability.

\subsection{Spontaneous symmetry breaking (SSB) of moving (tilted) solitons}

\begin{figure}[tph]
\begin{center}
\subfigure[]{\includegraphics[scale=0.5]{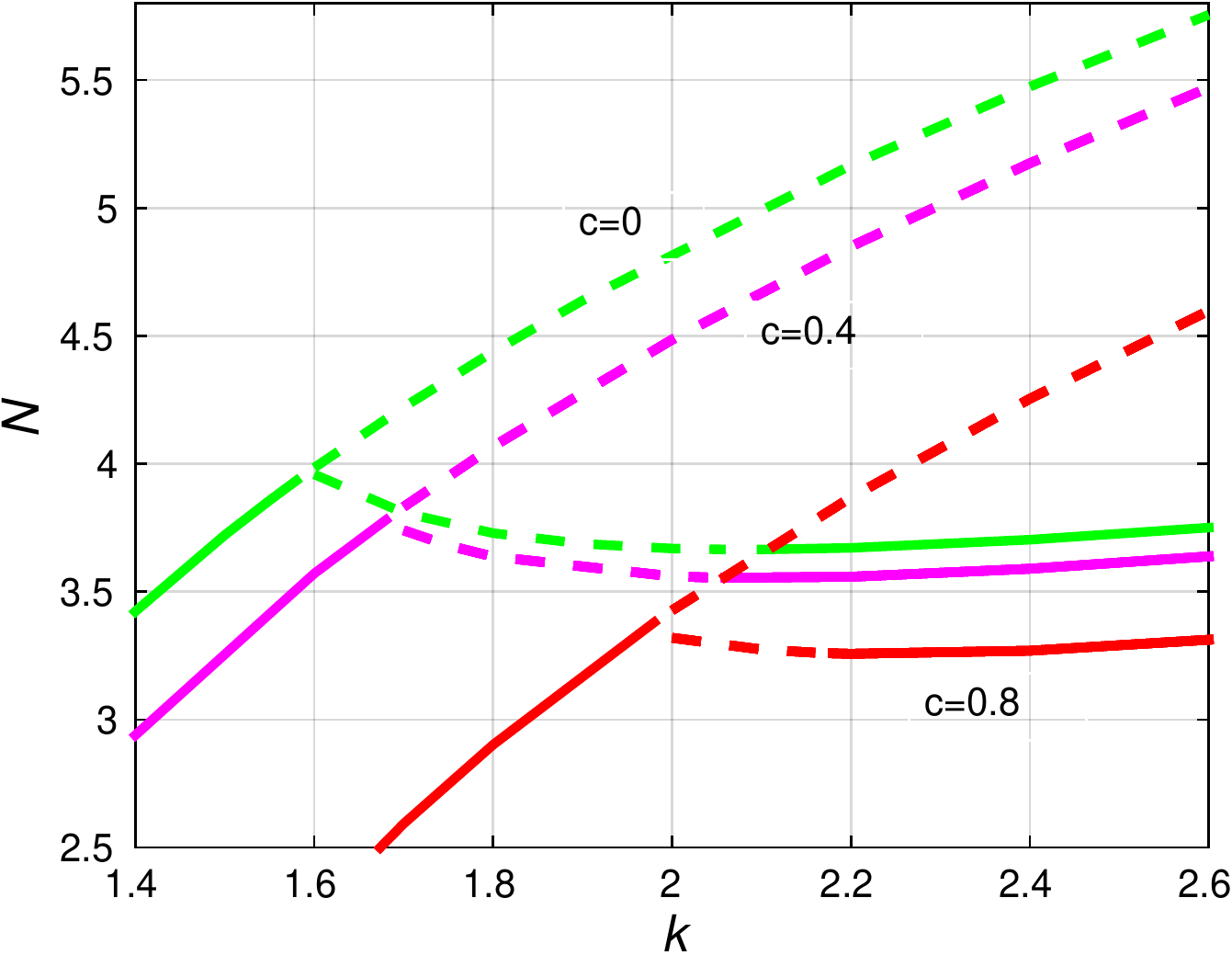}} %
\subfigure[]{\includegraphics[scale=0.5]{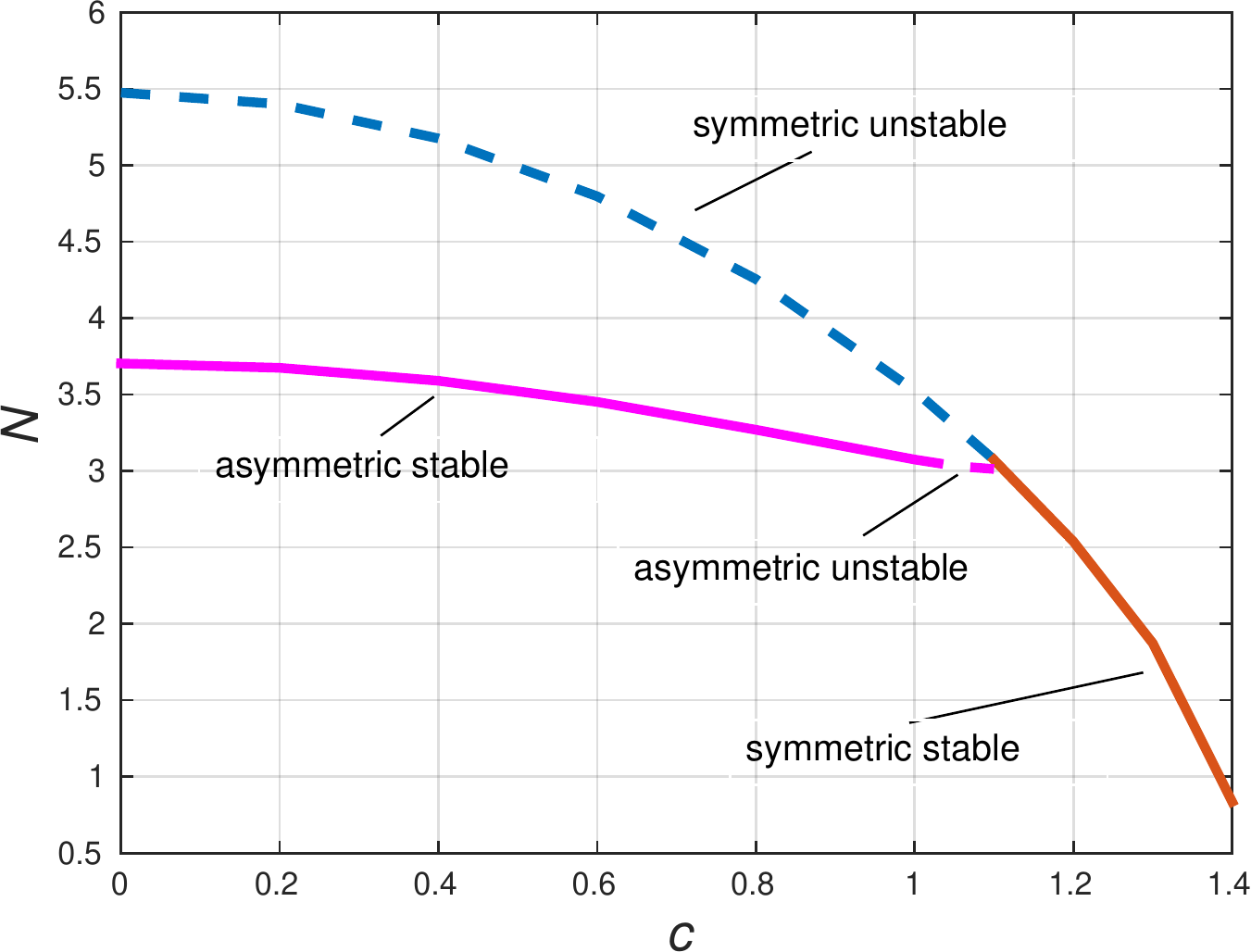}}
\end{center}
\caption{(a) SSB bifurcation diagrams at LI $\protect\alpha =1.6$ for
different values of the soliton's speed (tilt) $c$. (b) Power $N$ versus $c$
for fixed $\protect\alpha =1.6, k=2.4$.}
\label{bifur-c}
\end{figure}

As mentioned above, the incompatibility of the Galilean invariance with the
fractional diffraction makes velocity (tilt) $c$ of the solitons a
nontrivial parameter in Eq. (\ref{cc-stationary}). First, we address effects
of $c$ on the SSB bifurcation, which was not addressed in previous works.
For selected values of $c$, bifurcation diagrams produced by the numerical
solution of Eq. (\ref{cc-stationary}) are displayed in Fig.~\ref{bifur-c}%
(a), which demonstrates that the bifurcation keeps its subcritical
character. The shift of the $N(k)$ curves to larger $k$ is similar to the
effect of the Galilean boost in the case of the usual diffraction ($\alpha
=2 $): in that case, the removal of the velocity terms by means of the boost
produces a shift of the propagation constant $\Delta k=c^{2}/2$. For values $%
c=0.4$ and $0.8$, which are presented in Fig.~\ref{bifur-c}(a), this
expression yields $\Delta k(c=0.4)=0.08$ and $\Delta k(c=0.8)=0.32$, being
close to the shifts observed in Fig.~\ref{bifur-c}(a). A new effect, which
is absent in the case of the usual diffraction, is the decrease of norm $N_{%
\mathrm{SSB}}$ at the bifurcation point with the increase of $c$. It can be
explained by the fact that, for complex profile functions $U_{1,2}(x)$, in
the case of $c\neq 0$, the coupling Hamiltonian is $H_{\mathrm{coupling}%
}=\int_{-\infty }^{+\infty }$Re$\left\{ U_{1}(x)U_{2}^{\ast }(x)\right\} dx$%
, cf. Eq. (\ref{H}). Its value is reduced due to averaging of oscillations
of the complex integrand. In turn, the relative attenuation of the coupling
naturally leads to a decrease of value $N_{\mathrm{SSB}}$ at the point where
the nonlinearity becomes strong enough to initiate the SSB.

Another manifestation of the effect of $c$ on the families of stable and
unstable symmetric and asymmetric solitons is displayed in Fig.~\ref{bifur-c}%
(b) for fixed values of LI $\alpha $ and propagation constant $k$. It is
seen that tilt $c$ can be used to switch the optical beams between the
asymmetric and symmetric shapes, which may find applications in the design
of photonic devices. The trend towards the decrease of $N$ for large values
of $c$, especially for the symmetric solitons (for which the two-component
structure is not essential, while the fractionality remains a dominant
factor), can be explained by noting that rapid oscillations of the complex
wave function $\psi (x)$ caused by large $c$ make the value of the
fractional derivative in Eq. (\ref{FracDefi}) smaller. Therefore, the
strength of the self-focusing term (determined by the value of the norm),
which is necessary to balance the fractional diffraction in solitons,
becomes lower. 

\subsection{ Collisions between moving solitons}

Once stable solitons are produced in the moving (tilted) state, they can be
used to explore collisions of soliton pairs in the coupler \cite{Scripta}.
For this purpose, two solitons were numerically constructed as solutions of
Eq. (\ref{cc-stationary}), $u_{1,2}^{\pm }$, with velocities $\pm c$. Then,
direct simulations of Eq. (\ref{1D}) were run, with the input in the form of
the pair of solitons $u_{1,2}^{\pm }(x)$ placed, respectively, at $x<0$ and $%
x>0$, with a sufficiently large distance between them.

Here we focus on the following (most natural) settings for collisions
between mutually symmetric solitons, with equal values of the propagation
constant, $k$: (1) two stable symmetric solitons; (2) two stable asymmetric
solitons, with the same $k$, in the \textit{flipped} configuration, where
soliton $u_{1,2}^{+}$ has a larger component $u_{1}$ and a smaller one $%
u_{2} $, and vice versa for $u_{1,2}^{-}$ (cf. Ref. \cite{Scripta}); and (3)
two stable asymmetric solitons, which are mirror images of each other.

\begin{figure}[tph]
\centering
\begin{tabular}{cc}
\includegraphics[scale=0.6]{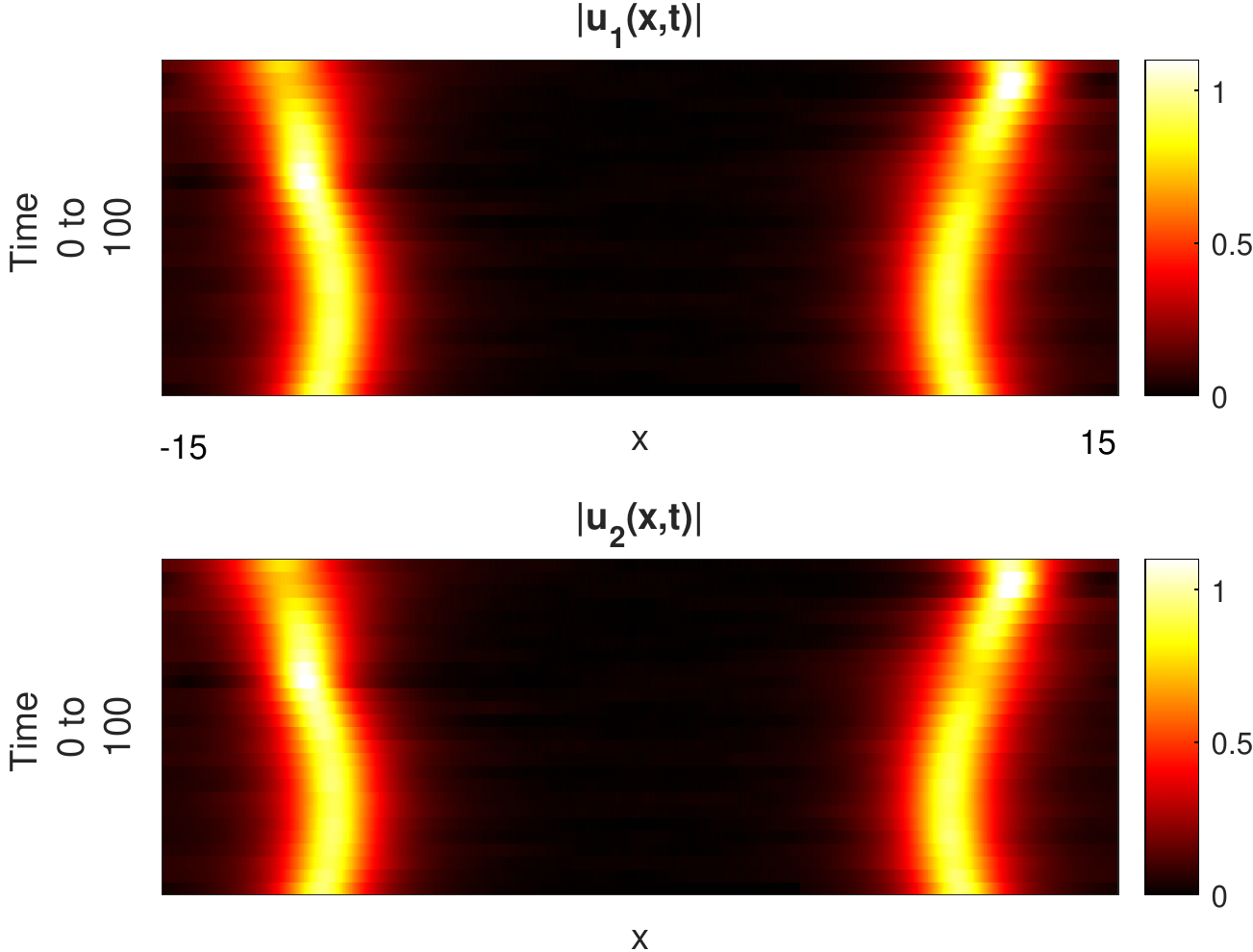} & %
\includegraphics[scale=0.6]{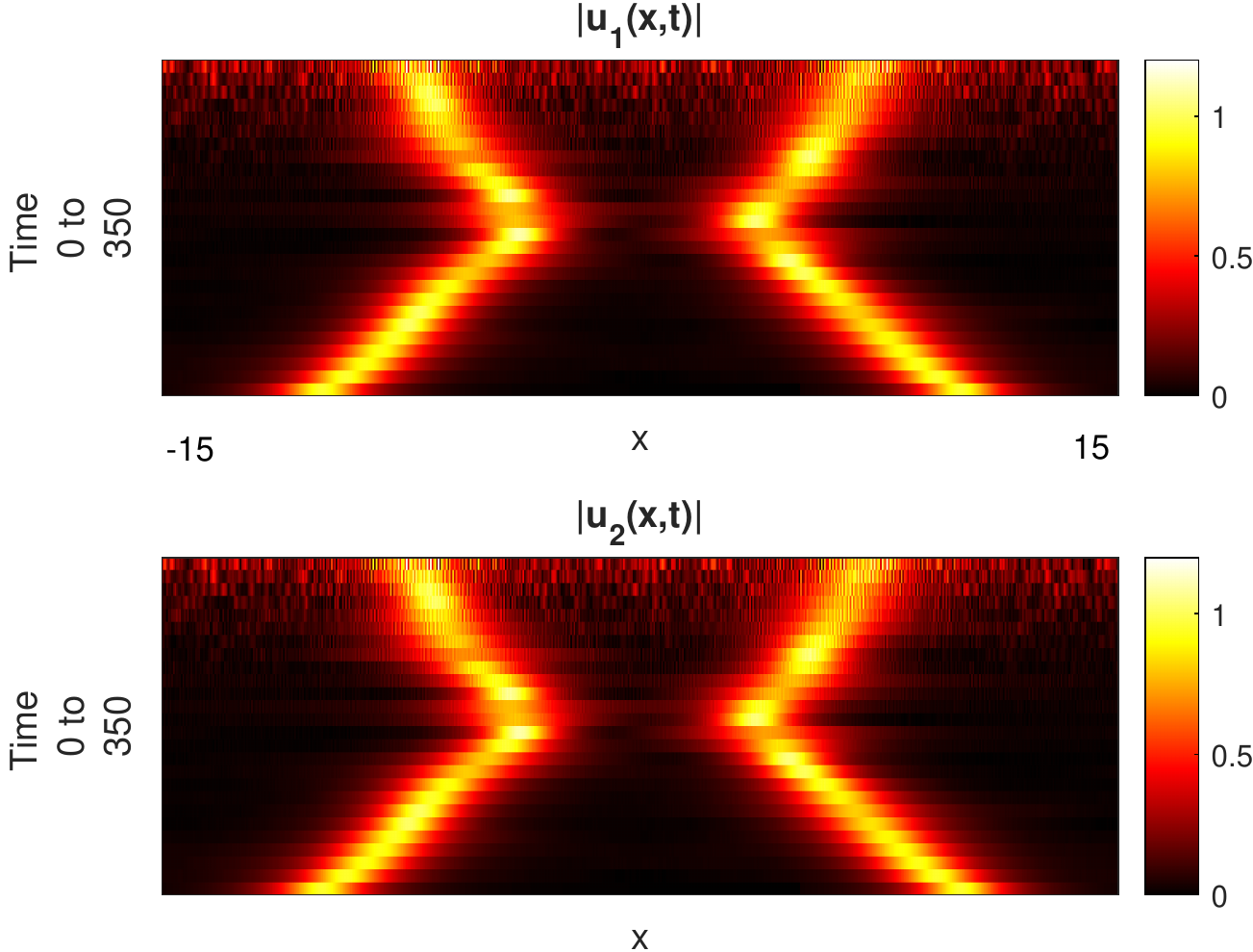} \\
$c=0.03$ & $c=0.06$ \\
&  \\
\includegraphics[scale=0.6]{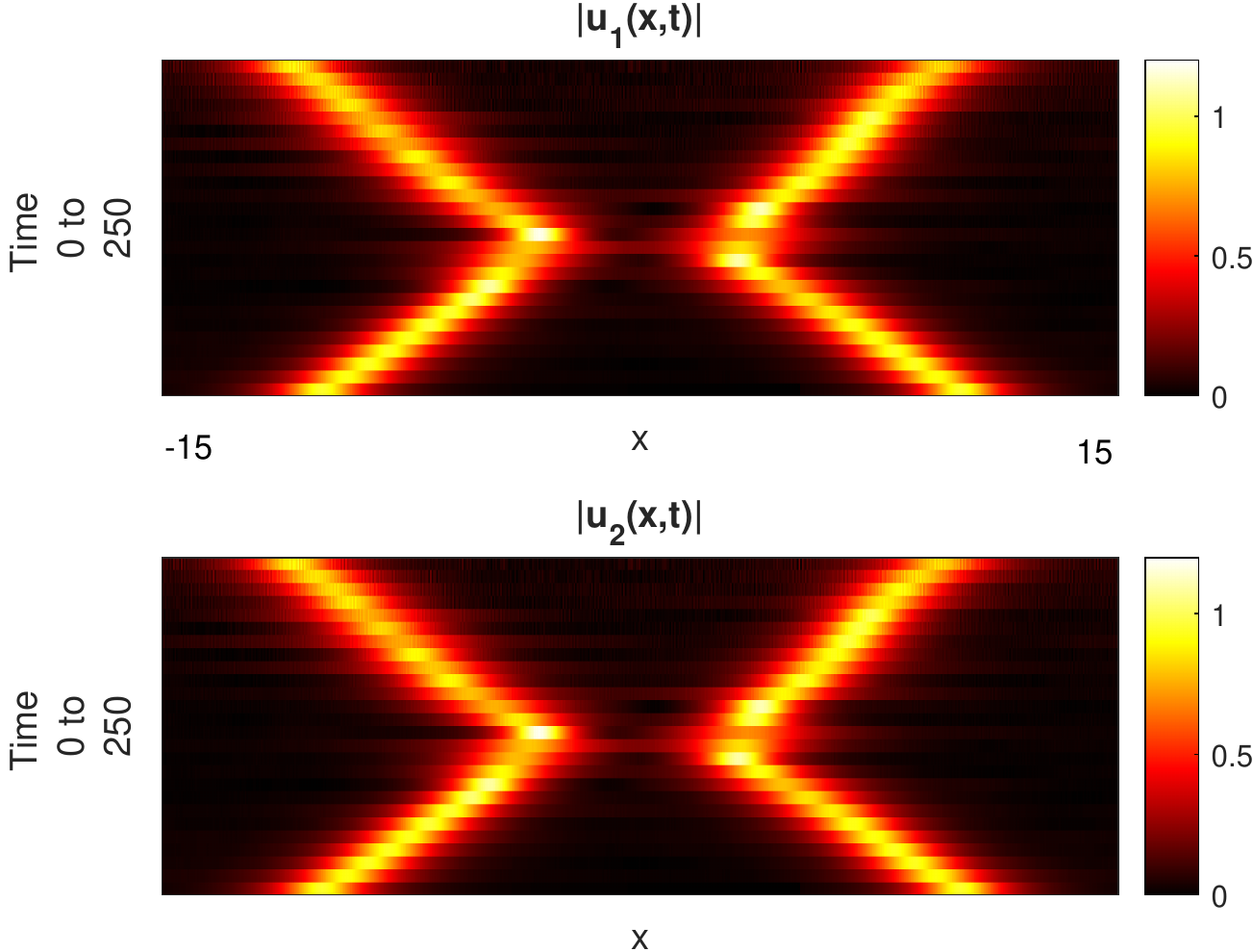} & %
\includegraphics[scale=0.6]{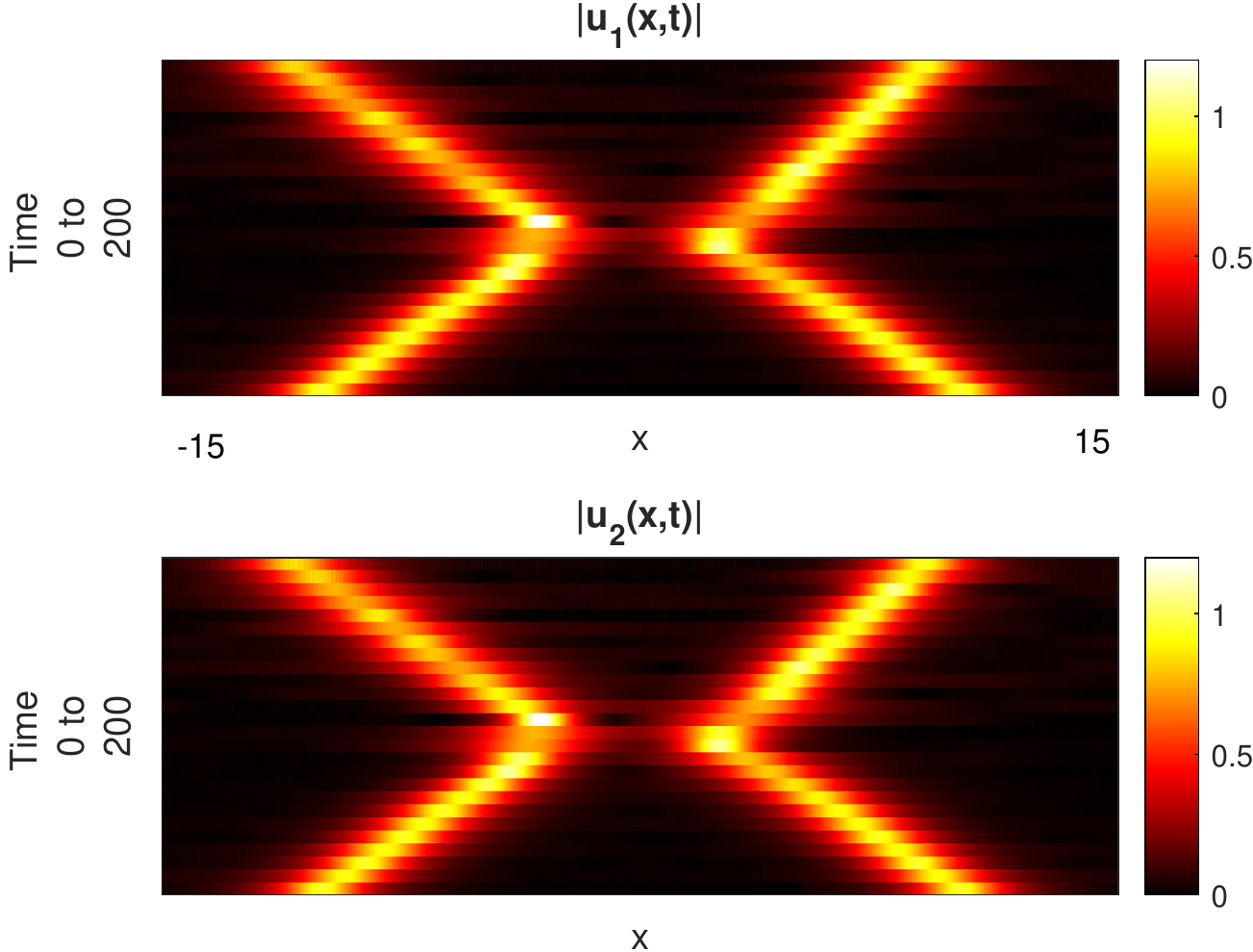} \\
$c=0.08$ & $c=0.09$ \\
&  \\
\includegraphics[scale=0.6]{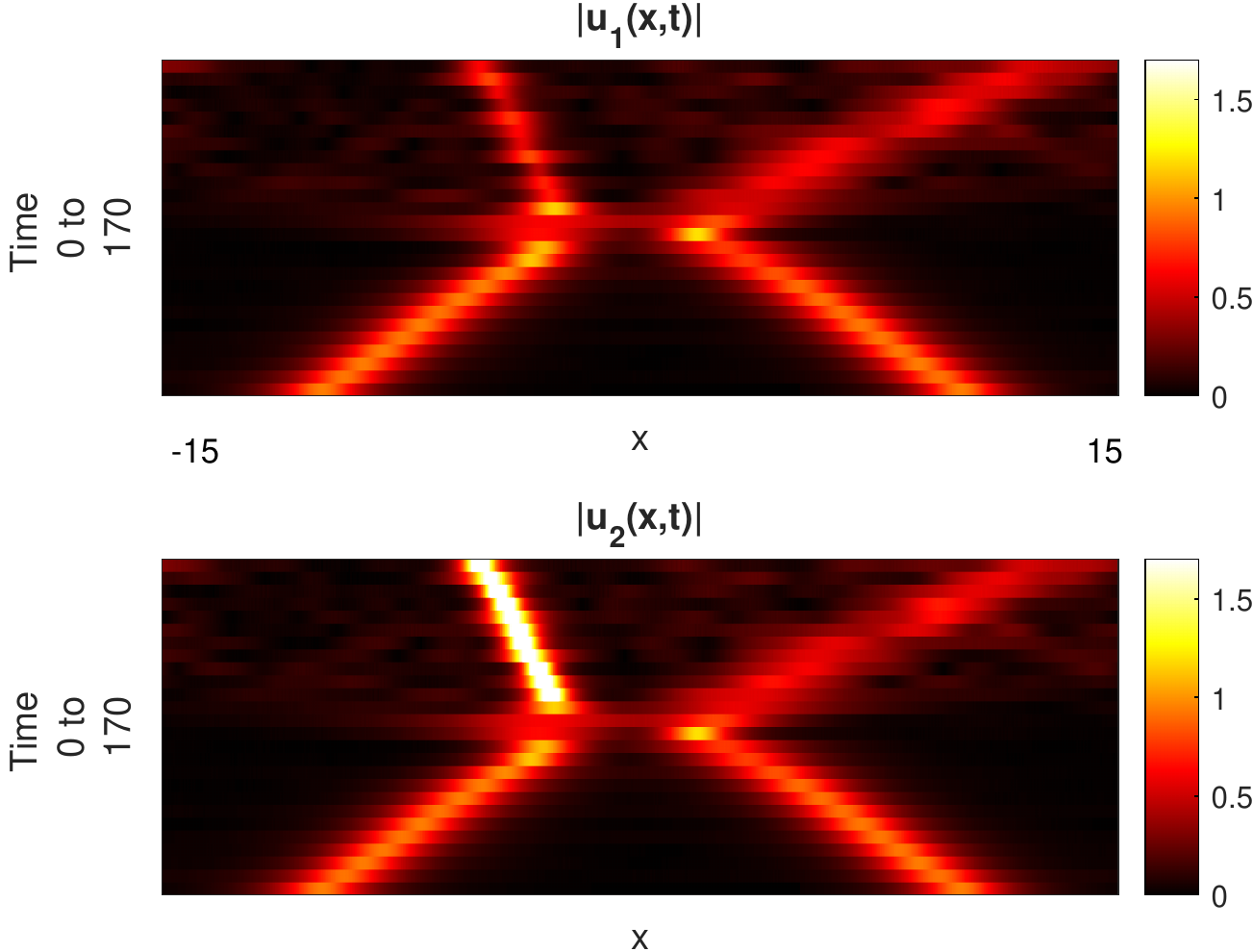} & %
\includegraphics[scale=0.6]{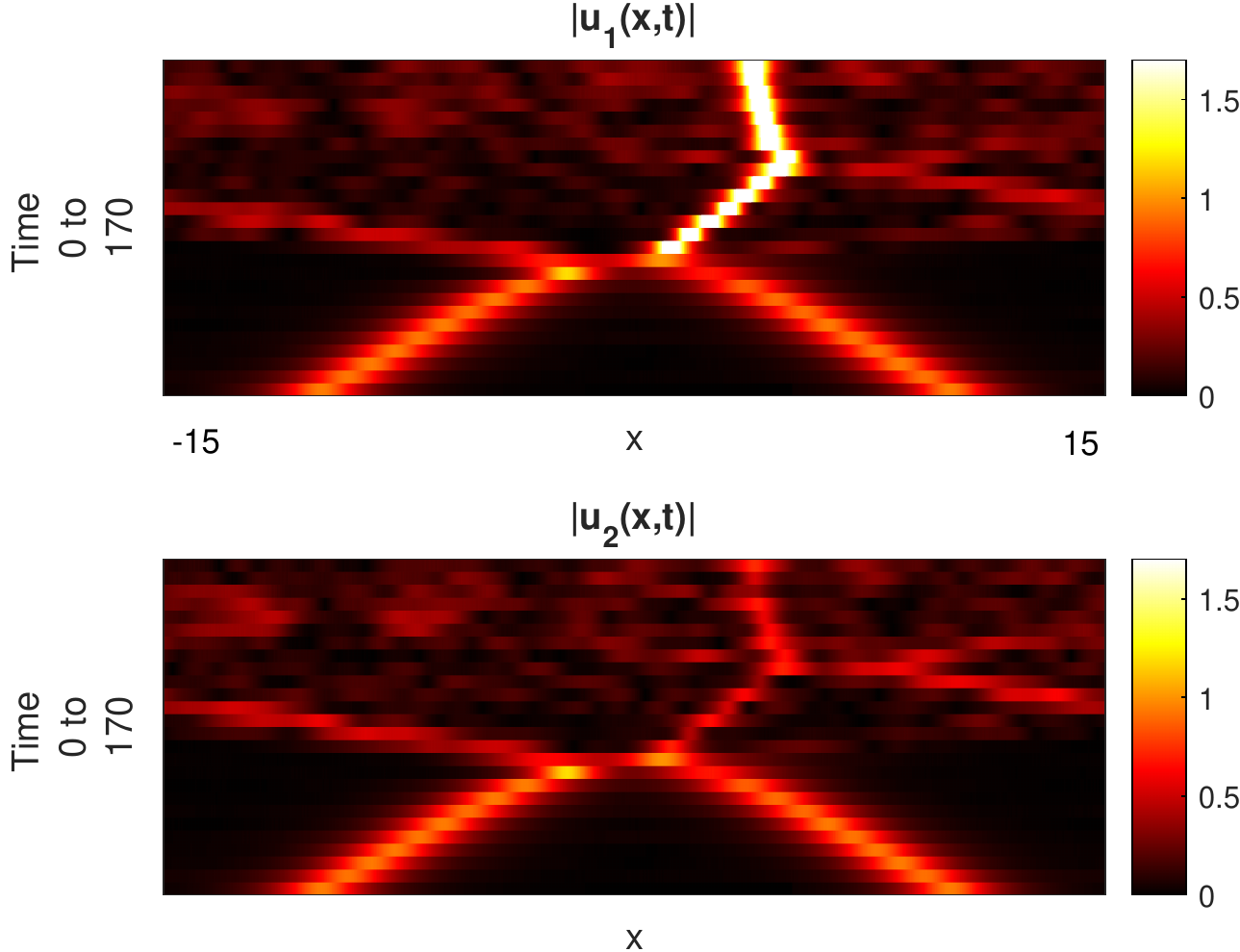} \\
$c=0.1$ & $c=0.12$ \\
&  \\
&
\end{tabular}%
\caption{The onset of SSB in collisions of slowly moving symmetric solitons,
with $\protect\alpha =1.6$, $k=1.4$, and velocities $\pm c$. The norms of
the solitons are $N=3.420$ ($c=0.03)$, $N=3.412$ ($c=0.06$), $N=3.404$ ($%
c=0.08$), $N=3.399$ ($c=0.09$), $N=3.393$ ($c=\pm 0.10$), $N=3.380$ ($c=\pm
0.12$).}
\label{col-sym-slow}
\end{figure}

Outcomes of collisions between stable symmetric solitons, at gradually
increasing speeds $\pm c$, are presented in Fig.~\ref{col-sym-slow}. In all
cases, the colliding solitons bounce back -- naturally, remaining far
separated for smallest speeds, and approaching closer to each other for
larger $c$. Up to the case of $c=0.06$, the entire picture remains fully
symmetric, with respect to both the two components in each soliton, and two
colliding solitons as well. Next, starting from $c=0.08$, the simulations
demonstrate onset of collision-induced SSB, which becomes obvious in the
case of $c=0.10$. In this case, the collision breaks the symmetry between
the components, as well as between the colliding solitons. In particular, it
is worthy to note that the post-collision amplitude of component $u_{2}$ in
the left soliton is much larger than before the collision. The
collision-induced SSB effect is explained by the instability of the
transient state formed by the colliding solitons when they are separated by
a relatively small distance. A qualitatively similar SSB mechanism was
discovered by simulations of soliton-soliton collisions in the single
equation with the normal diffraction ($\alpha =2$) and cubic-quintic
nonlinearity \cite{Khayk}.

With subsequent growth of $c$, the collision picture remains approximately
the same as shown in Fig. \ref{col-sym-slow} for $c=0.12$ up to $c=0.2$ (not
shown here in detail). At still larger speeds, the symmetry is gradually
recovered, leading, eventually, to practically elastic collisions at $c\geq
0.34$, which is a natural outcome of fast collisions \cite{RMP}.
%

Results of collisions between stable asymmetric solitons in the mutually
flipped states, as defined above, are demonstrated in Fig. \ref%
{col-flip-slow}. The general picture is similar to that outlined above for
the collisions between symmetric solitons. Namely, at low speeds, $c\leq
0.04 $, the solitons bounce back, without breaking the symmetry between the
colliding ones. In fact, in this case each soliton switches from the
intrinsic asymmetric shape into a nearly symmetric one, as concerns the
relation between its two components. Then, starting from $c=0.06$, the
collision-induced SSB\ effect sets in, leading to strong symmetry breaking
at $c=0.1$, with a dominant $u_{1}$ component of the left soliton in the
post-collision state. Approximately the same inelastic outcome of the
collision persists up to $c\simeq 0.40$ (not shown here in detail), while
the further increase of the speed gradually leads to a transition to
quasi-elastic collisions.

\begin{figure}[tph]
\centering
\begin{tabular}{cc}
\includegraphics[scale=0.6]{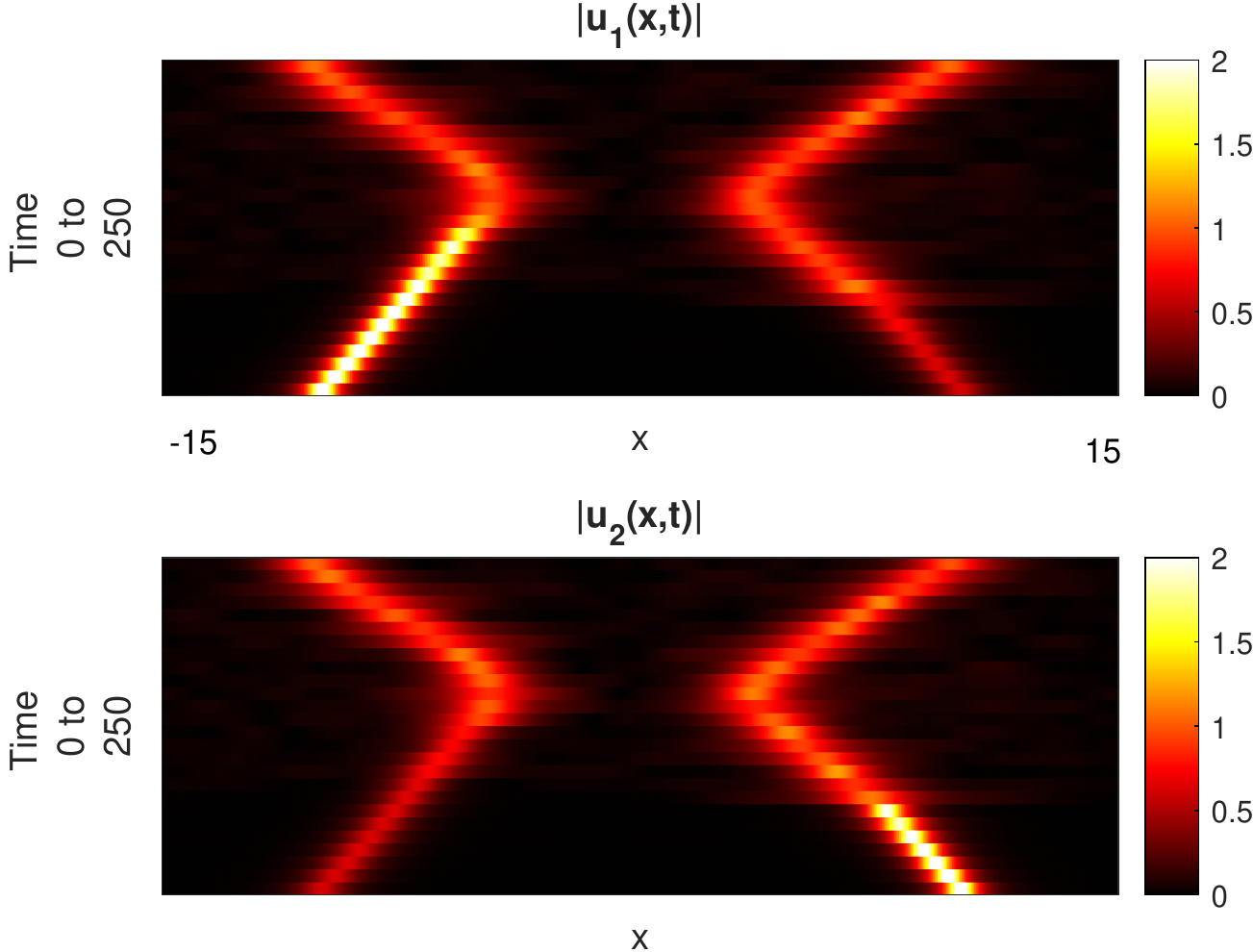} & %
\includegraphics[scale=0.6]{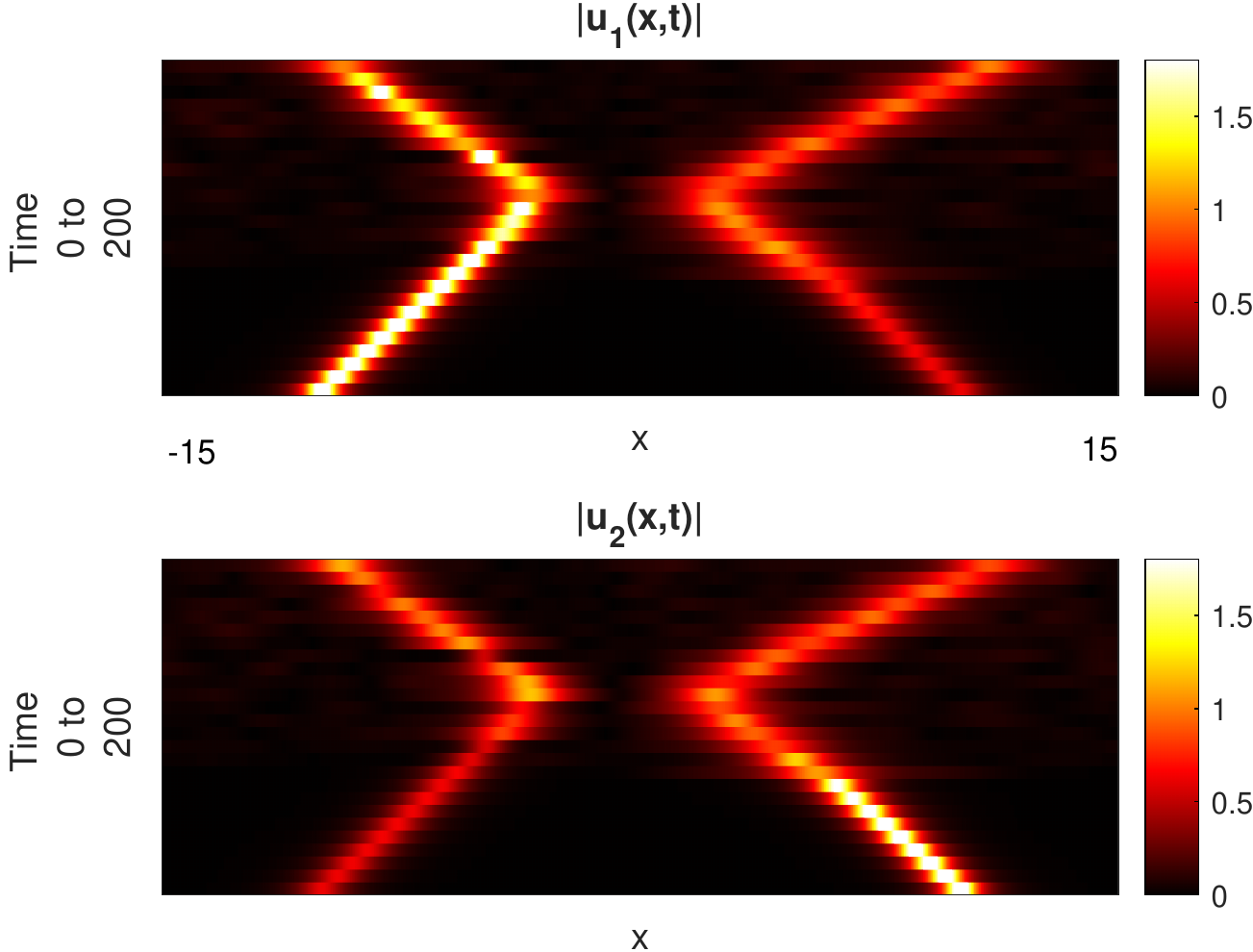} \\
$c=0.04$ & $c=0.06$ \\
\includegraphics[scale=0.6]{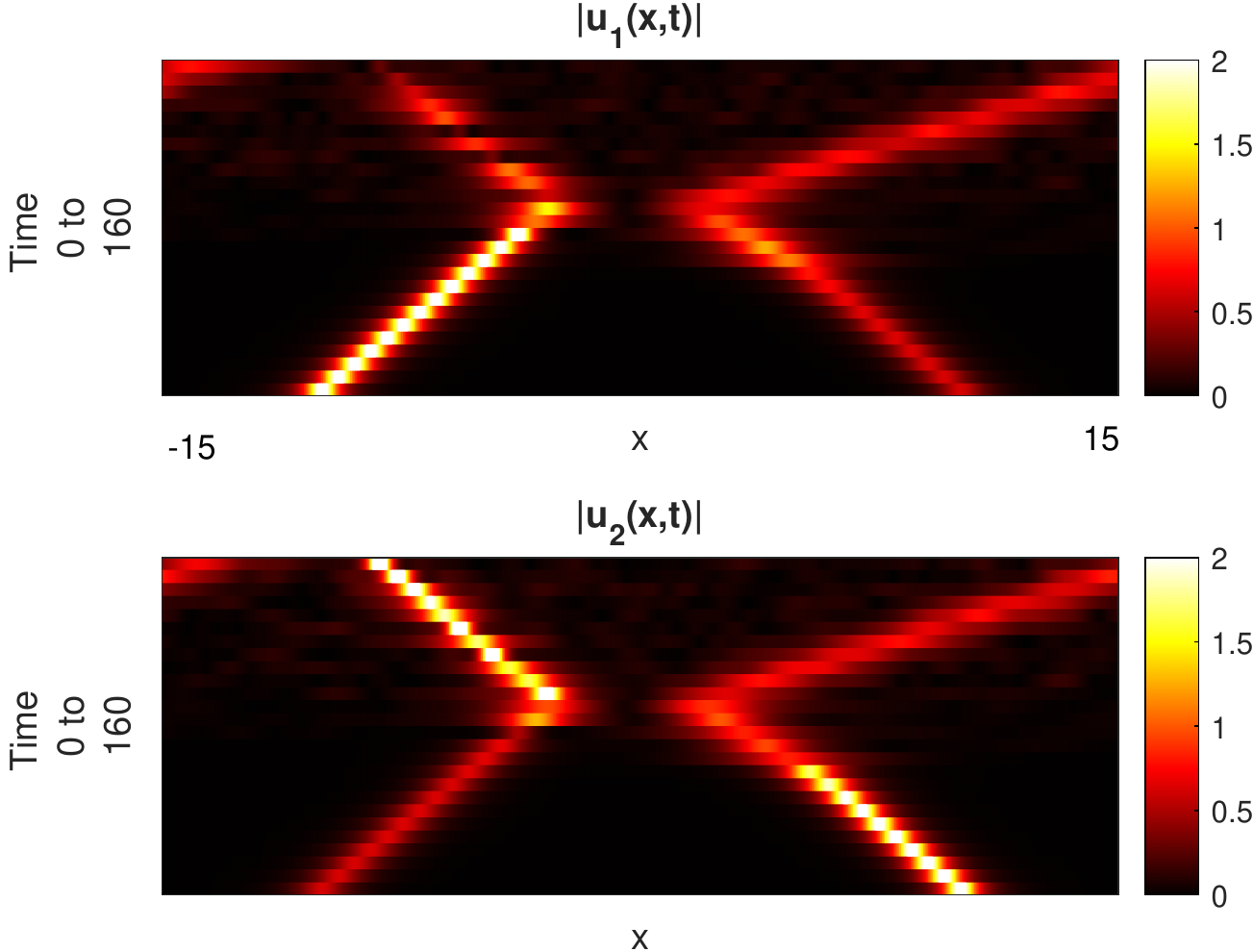} & %
\includegraphics[scale=0.6]{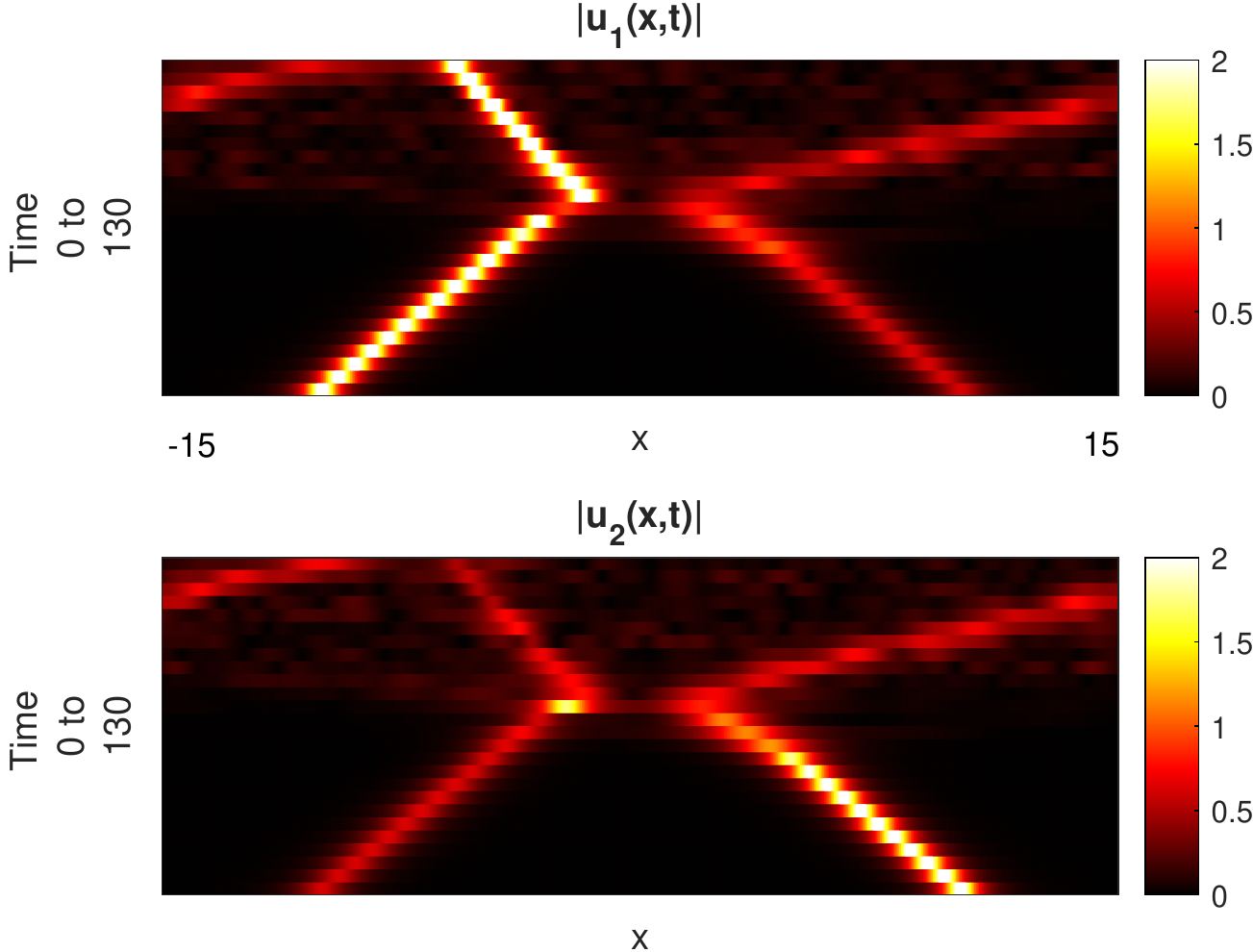} \\
$c=0.08$ & $c=0.10$%
\end{tabular}%
\caption{The gradual onset of SSB in collisions of slowly moving mutually
flipped asymmetric solitons, with $\protect\alpha =1.6$, $k=2.6$, and
velocities $\pm c$. The norms of the solitons are $N=3.749$ ($c=0.04$), $%
N=3.748$ ($c=0.06$), $N=3.746$ ($c=0.08$), $N=3.744$ ($c=0.1$), cf. Fig.
\protect\ref{col-sym-slow}.}
\label{col-flip-slow}
\end{figure}

\begin{figure}[tph]
\subfigure[]{%
\includegraphics[scale=0.6]{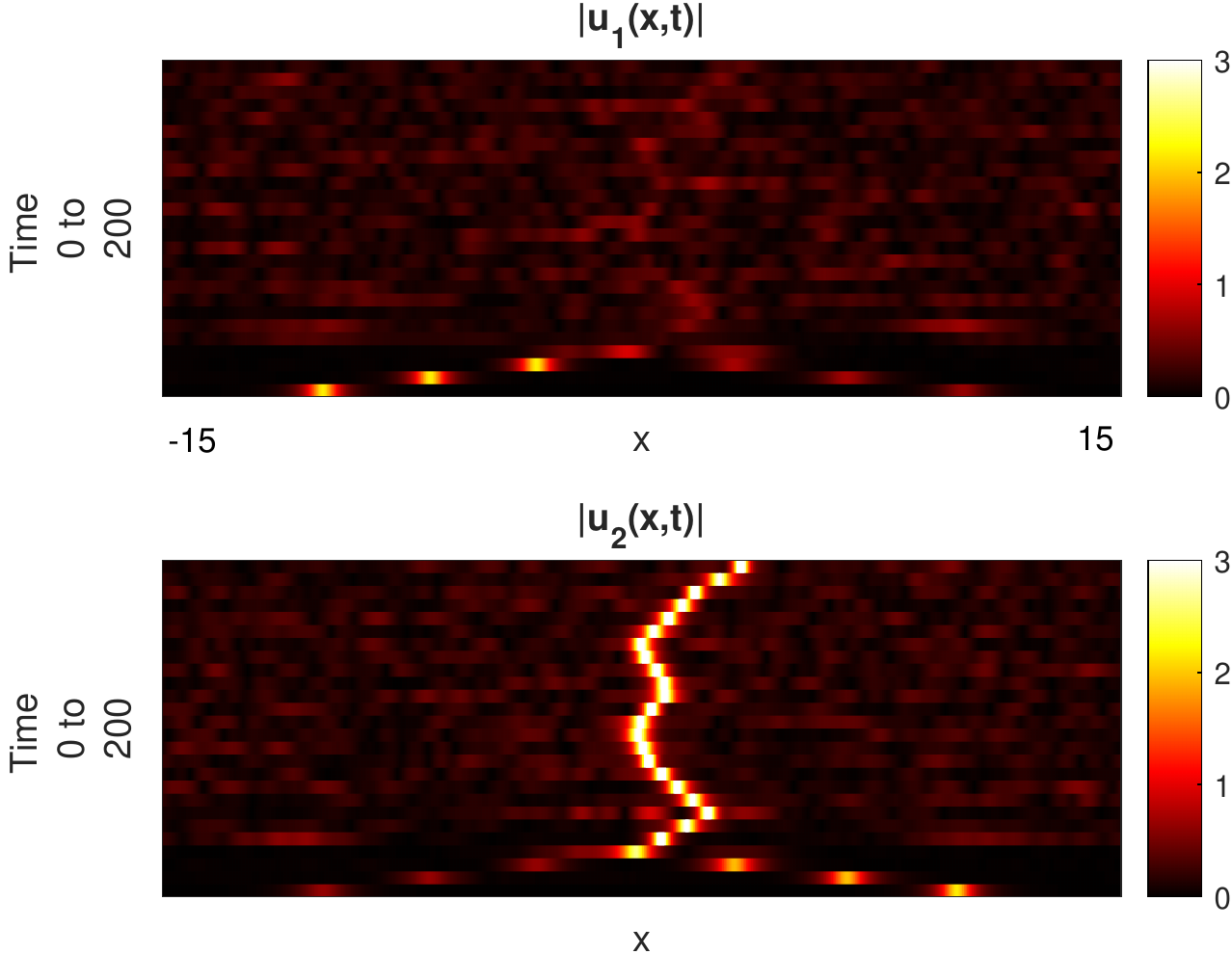}}
\subfigure[]{%
\includegraphics[scale=0.6]{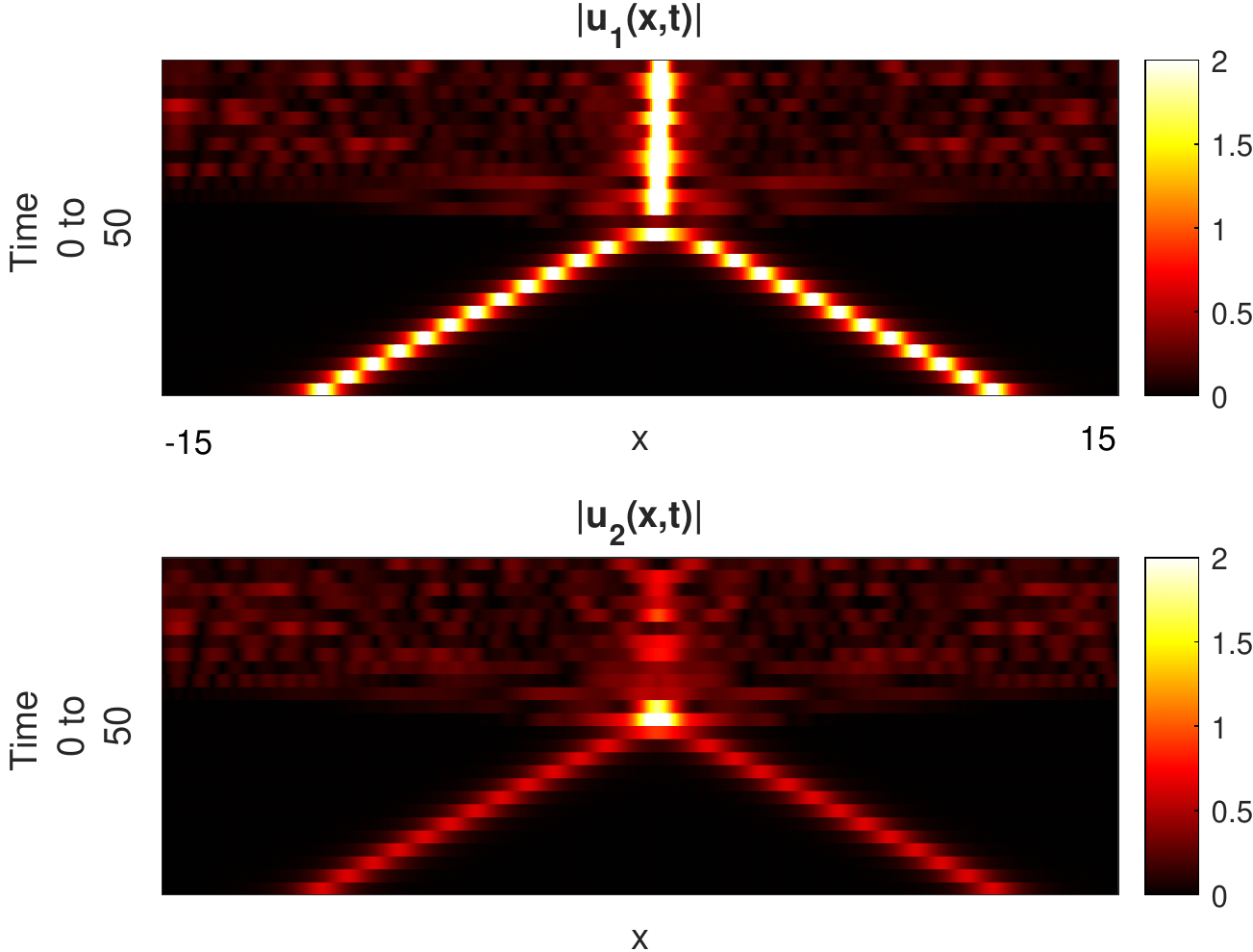}}
\caption{Examples of strongly inelastic collisions of solitons with $k=2.6$
at $\protect\alpha =1.6$. (a)The post-collision dynamics of originally
flipped asymmetric solitons for $c=0.42$ and $N=3.627$. (b) Merger of
asymmetric solitons colliding in the unflipped state, for $c=0.4$ and $%
N=3.638$.}
\label{col-flip-long}
\end{figure}

The situation when a fully inelastic collision of the mutually flipped
asymmetric solitons gives rise to very strong symmetry breaking is
additionally illustrated by Fig. \ref{col-flip-long}(a), for the same values
of $\alpha $ and $k$ as in Fig. \ref{col-flip-slow}, and $c=0.42$. It is
observed that, in the post-collision state, component $u_{1}$ almost
vanishes, while nearly all the initial norm is trapped in component $u_{2}$
in the form of a quasi-soliton, which performs slow erratic motion.
Actually, the dominant component and direction of motion of the emerging
mode are selected by the system randomly (as confirmed by additional
numerical results), as a result of the above-mentioned instability in the
transient state created by the collision.

Lastly, in Fig.~\ref{col-flip-long}(b) we present an example of a fully
inelastic collision between identical stable asymmetric solitons, i.e., ones
in the unflipped configuration. In this case, the solitons merge into a
single strongly asymmetric quiescent one, with the same dominant component $%
u_{1}$ as in the original solitons.

\section{Conclusion}

As a contribution to the quickly developing studies of solitons in systems
with fractional diffraction, we have addressed the phenomenology of the SSB
(spontaneous symmetry breaking) in the one-dimensional dual-core system,
with the Riesz fractional derivative and cubic self-focusing acting in the
cores, and linear coupling between the cores. The corresponding system of
FNLSEs (fractional nonlinear Schr\"{o}dinger equations) models
tunnel-coupled planar optical waveguides with the fractional diffraction, as
well as coupled waveguides with the fractional group-velocity dispersion in
the temporal domain (the latter setting was recently realized in the
experiment \cite{Shilong}, and may be appropriate for the realization of the
results predicted in the present work).

By means of systematic numerical computations and the analytical method
based on the VA (variational approximation), we have identified the SSB in
the system as the bifurcation of the subcritical type, i.e., the phase
transition of the first kind. With the increase of the fractionality, i.e.,
parameter $2-\alpha $, where $\alpha $ is the LI (L\'{e}vy index), the
subcritical character of the SSB bifurcations gets stronger pronounced, in
comparison with very weak subcriticality in the case of the usual
(non-fractional) diffraction, $\alpha =2$. In the limit of $\alpha
\rightarrow 1$, the bifurcation becomes the \textit{extreme subcritical}
one, i.e., with the backward-going branches of asymmetric solitons which
never turn forward. The VA produces reasonably accurate results for the SSB,
in spite of the complex structure of the system. The (in)stability of
families of asymmetric solitons exactly follows the Vakhitov-Kolokolov
criterion. By means of the numerical method, the study of the SSB
bifurcation and families of symmetric and asymmetric solitons has been
extended for moving (tilted) ones, which is a nontrivial issue for the
fractional system, as it breaks the Galilean invariance. Collisions between
moving solitons are systematically studied too, demonstrating another SSB
effect, as well as merger of the colliding solitons.

As an extension of the analysis, it may be relevant to consider a similar
system of linearly coupled fractional complex Ginzburg-Landau equations,
with the objective to predict symmetric and asymmetric dissipative solitons,
as well as bound states of such solitons in the respective laser systems,
cf. Ref. \cite{Sigler} where a similar analysis was developed in the case of
the non-fractional diffraction.

\section*{Acknowledgment}

We appreciate a useful discussion with W. B. Cardoso. The work of B.A.M. was
supported, in part, by grant No. 1695/22 of the Israel Science Foundation.

\end{document}